\documentclass[fleqn,usenatbib,useAMS]{mnras}

\usepackage{graphicx}	% Including figure files
\usepackage{amsmath}	% Advanced maths commands
\usepackage{amssymb}	% Extra maths symbols
\usepackage{multicol}        % Multi-column entries in tables
\usepackage{bm}		% Bold maths symbols, including upright Greek
\usepackage{pdflscape}	% Landscape pages

\usepackage[T1]{fontenc}
\usepackage{ae,aecompl}

\title[The ALHAMBRA-Survey $K_s$-band catalogue]{A $K_s$-band selected catalogue of objects in the ALHAMBRA survey}

\author[L. Nieves-Seoane]{L. Nieves-Seoane$^{1,2,3}$
\thanks{Contact e-mail: \href{mailto:nievesl@ifca.unican.es}{nievesl@ifca.unican.es}}, 
    A.~Fernandez-Soto$^{1,4}$,
	P.~Arnalte-Mur$^{2,3}$,
	A.~Molino$^{5}$,
\newauthor
	M.~Stefanon$^{6}$,
    I.~Ferreras$^{7}$,
	B.~Ascaso$^{8}$,
	F.J.~Ballesteros$^{2}$,
	D.~Crist\'obal-Hornillos.$^{9}$,
\newauthor
    C.~L\'opez-Sanju\'an$^{9}$,
    Ll.~Hurtado-Gil$^{1,2}$,
    I.~M\'arquez$^{12}$,
    J.~Masegosa$^{12}$,
\newauthor
	J.A.L.~Aguerri$^{10,11}$,
	E.~Alfaro$^{12}$,
	T.~Aparicio-Villegas$^{13}$,
	N.~Ben\'{\i}tez$^{12}$,
\newauthor
	T.~Broadhurst$^{14,15}$,
	J.~Cabrera-Ca\~no$^{16}$,
	F.J.~Castander$^{17}$,
	J.~Cepa$^{10,11}$,
	M.~Cervi\~no$^{10,11,12}$,
\newauthor
	R.M.~Gonz\'alez~Delgado$^{12}$,
	C.~Husillos$^{12}$,
	L.~Infante$^{18}$,
	V.~J.~Mart\'{\i}nez$^{2,3,4}$,
	M.~Moles$^{9,12}$,
\newauthor
	A.~del~Olmo$^{12}$,
	J.~Perea$^{12}$,
	M.~Povi\'c$^{12}$,
	F.~Prada$^{12,19,20}$,
	J.M.~Quintana$^{12}$,
\newauthor
    P.~Troncoso-Iribarren$^{18}$,
	K.~Viironen$^{9}$
\\
$^{1}$Instituto de F\'{\'i}sica de Cantabria (CSIC-UC). Av. de los Castros s/n, E-39005 Santander (SPAIN)\\
$^{2}$Observatori Astron\`omic de la Universitat de Val\`encia. C/ Cat. Jos\'e Beltr\'an 2, E-46980 Paterna (SPAIN)\\
$^{3}$Departament d'Astronomia i Astrof\'isica, Universitat de Val\`encia, E-46100, Burjassot (SPAIN)\\
$^{4}$Unidad Asociada Observatorio Astron\'omico (CSIC-UV). C/ Cat. Jos\'e Beltr\'an 2, E-46980 Paterna (SPAIN)\\
$^{5}$Instituto de Astronom{\'{\i}}a, Geof{\'{\i}}sica e Ci\'encias Atmosf\'ericas, Universidade de S{\~{a}}o Paulo, S{\~{a}}o Paulo (BRAZIL)\\
$^{6}$Leiden Observatory, Leiden University, PO Box 9513, NL-2300 RA Leiden (THE NETHERLANDS)\\
$^{7}$Mullard Space Science Laboratory, University College London, Holmbury St. Mary, Dorking, Surrey RH5 6NT (UK)\\
$^{8}$APC, AstroParticule et Cosmologie, Universit\'e Paris Diderot, CNRS/IN2P3, CEA/lrfu, Observatoire de Paris, \\Sorbonne Paris Cit\'e, 10, rue Alice Domon et L\'eonie Duquet, 75205 Paris Cedex 13 (FRANCE)\\
$^{9}$Centro de Estudios de F\'isica del Cosmos de Arag\'on, Plaza San Juan 1, E-44001 Teruel (SPAIN)\\
$^{10}$Instituto de Astrof\'isica de Canarias, V\'ia L\'actea s/n, E-38200 La Laguna, Tenerife (SPAIN)\\
$^{11}$Departamento de Astrof\'isica, Facultad de F\'isica, Universidad de La Laguna, E-38206 La Laguna (SPAIN)\\
$^{12}$IAA-CSIC, Glorieta de la Astronom\'ia s/n, E-18008 Granada (SPAIN)\\
$^{13}$Observat\'orio Nacional-MCT, Rua Jos\'e Cristino, 77. CEP 20921-400, Rio de Janeiro-RJ (BRAZIL)\\
$^{14}$Department of Theoretical Physics, University of the Basque Country UPV/EHU, E-48080 Bilbao (SPAIN)\\
$^{15}$IKERBASQUE, Basque Foundation for Science, Bilbao (SPAIN)\\
$^{16}$Departamento de F\'isica At\'omica, Molecular y Nuclear, Facultad de F\'isica, Universidad de Sevilla, E-41012 Sevilla (SPAIN)\\
$^{17}$Institut de Ci\`encies de l'Espai (IEEC-CSIC), Facultat de Ci\`encies, Campus UAB, E-08193 Bellaterra (SPAIN)\\
$^{18}$Departamento de Astronom\'ia, Pontificia Universidad Cat\'olica. 782-0436 Santiago (CHILE)\\
$^{19}$Instituto de F\'{\i}sica Te\'orica, (UAM/CSIC), Universidad Aut\'onoma de Madrid, Cantoblanco, E-28049 Madrid (SPAIN)\\
$^{20}$Campus of International Excellence UAM+CSIC, Cantoblanco, E-28049 Madrid (SPAIN) }

\date{Last updated 2000 January 1; in original form 2000 January 1}

\pubyear{2016}

\begin{document}
\label{firstpage}
\pagerange{\pageref{firstpage}--\pageref{lastpage}}
\maketitle

\begin{abstract}
 The original ALHAMBRA catalogue contained over 400,000 galaxies selected using a synthetic F814W image, to the magnitude limit AB(F814W)$\approx$24.5. Given the photometric redshift depth of the ALHAMBRA multiband data (<z>=0.86) and the approximately $I$-band selection, there is a noticeable bias against red objects at moderate redshift.
We avoid this bias by creating a new catalogue selected in the $K_s$ band. This newly obtained catalogue is certainly shallower in terms of apparent magnitude, but deeper in terms of redshift, with a significant population of red objects at $z>1$.
We select objects using the $K_s$ band images, which reach an approximate AB magnitude limit $K_s \approx 22$. We generate masks and derive completeness functions to characterize the sample. We have tested the quality of the photometry and photometric redshifts using both internal and external checks. 
Our final catalogue includes $\approx 95,000$ sources down to $K_s \approx 22$, with a significant tail towards high redshift. We have checked that there is a large sample of objects with spectral energy distributions that correspond to that of massive, passively evolving galaxies at $z > 1$, reaching as far as $z \approx 2.5$. We have tested the possibility of combining our data with deep infrared observations at longer wavelengths, particularly Spitzer IRAC data.\end{abstract}


%%%%%%%%%%%%%%%%%%%%%%%%%%%%%%%%%%%%%%%%%%%%%%%%%%

%%%%%%%%%%%%%%%%% BODY OF PAPER %%%%%%%%%%%%%%%%%%

\section{Introduction}

Astronomical surveys are one of the key elements in the advancement of our knowledge of celestial objects. From the earliest times astronomers have charted stars and observed their basic properties, namely their positions and apparent brightnesses. This task increased exponentially in complexity over the last centuries with the successive arrivals of the telescope, the photographic plate and the electronic detector. 

In our time some of the most successful astronomical surveys have aimed at covering ever larger fractions of the phase space that includes area in the sky, photometric depth and spectral information. For the moment being (and in any foreseeable future) no project will cover satisfactorily and simultaneously all of those dimensions. For example, the Sloan Digital Sky Survey \citep[SDSS,][]{2000AJ....120.1579Y} and the Two Degree Field Galaxy Redshift Survey \citep[2dFGRS,][]{2001MNRAS.328.1039C} have obtained spectral information for $\sim 10^5 \mbox{--} 10^6$ objects each, by observing large areas (approximately 1/4 of the whole sky) down to a relatively shallow limit (apparent magnitudes $AB \approx 19$). Their photometric counterparts cover areas in the sky of the same size, but reach ten times deeper, out to a typical magnitude $AB \approx 21 \mbox{--} 22$. At the other end of survey space, deep surveys like the Hubble Deep Fields \citep{2000ARA&A..38..667F} cover tiny areas of the sky (of the order of $10^{-3}$ square degrees or even less) but do include spectroscopy out to $AB \approx 25 \mbox{--} 26$ and multi-band photometry out to $AB \approx 28$ and even deeper.

A different "axis" defining cosmic surveys is that of spectral completeness. In the most basic end, early surveys like the Palomar Observatory Sky Survey  \citep[POSS, ][]{1963PASP...75..488M,1991PASP..103..661R}, included only photometric information in two different bands (\textit{i.e.} one colour) for each object. In the opposite end, spectroscopic surveys include a full spectrum for each target, with all that implies in terms of information content regarding measurements of redshift, star formation history, mass, metallicity, etc. Since the advent of the Hubble Deep Fields \citep{2000ARA&A..38..667F} and other surveys at the end of the last century it has become commonplace to obtain images through multiple filters both in the optical and the near infrared in order to measure at least some spectral properties of the targets, which should allow for basic estimation of some of the physical quantities that would otherwise need a full spectral analysis. The use of photometric redshift techniques has grown and become standard based on this kind of studies \citep{1999ApJ...513...34F,2000ApJ...536..571B,2000A&A...363..476B}. Over the last few years some surveys have been explicitly designed having these techniques in mind (COMBO-17, \citealt{2003A&A...401...73W}; ALHAMBRA, \citealt{moles2008}) and have proved the case for even larger surveys with multiple medium-band filter images (J-PAS, \citealt{2014arXiv1403.5237B}).

Early-type galaxies dominate the bright end of the luminosity function at low and moderate redshifts \citep{1997ApJ...475..494L}, in particular they include the most massive galaxies that inhabit the largest overdensities in those epochs. They represent the most massive and evolved objects in the second half of the life of the Universe, and their study is basic to understand how star formation proceeded and its interrelations with many other cosmic processes: black hole formation and evolution, galaxy clustering and the formation of large-scale structures, galactic interactions and mergers, and the AGN phenomenon \citep[][ and references therein]{2014ARA&A..52..589H}. Due to their intrinsically red colours, early-type galaxies are selected against in magnitude-limited surveys selected at optical wavelengths at all those redshifts where the Balmer break and associated absorption features around $\lambda = 4000$ \AA\ are redshifted into the detection band and redwards of it. Over the last years the development of several surveys that detect objects in near infrared (NIR) bands has significantly helped in the analysis of the evolution of early-type galaxies at moderate and high redshift, e.g. the Newfirm Medium Band Survey \citep[NMBS, ][]{2011ApJ...735...86W}, UKIDSS-Ultra Deep Survey \citep{2007MNRAS.379.1599L}, WIRCam Deep Survey \citep[WIRCDS, ][]{2012A&A...545A..23B}, and Ultra VISTA \citep{2012A&A...544A.156M,2013ApJS..206....8M}.

\begin{table}
\centering
\caption{Comparison with other photometric $K$-band selected surveys.}\label{surveycomp}
\begin{tabular}{lcc}
\hline
Survey  & Area 	&	 AB Magnitude	\\ 
 &  &  ($5\sigma$ Limit) \\ 
\hline
MUSYC& 0.015 $\mathrm{deg^2}$ & $K_s \approx 22.5 $ \\ 
NMBS& 0.44 $\mathrm{deg^2}$ & $K \approx 24.2 $\\ 
UKIDSS-UDS& 0.77 $\mathrm{deg^2}$ & $K \approx 24.6 $\\ 
WIRCDS &  2.03 $\mathrm{deg^2}$& $K_s\approx 24.0 $\\ 
UVISTA  &  1.50 $\mathrm{deg^2}$ & $K_s\approx 23.8$ \\ 
ALHAMBRA $K_s$-band  &  2.47 $\mathrm{deg^2}$  & $K_s \approx 21.5$\\ 
\hline
\end{tabular}
\end{table}

In the particular case of the ALHAMBRA survey, where detection is performed over a synthetic image that emulates the Hubble Space Telescope F814W filter, this selection effect that creates a bias against red galaxies begins to be noticeable at $z\approx 0.8$, and is dominant at $z\geq 1.1$, as has already been noticed by \citet{2014MNRAS.441.1783A}. A typical early-type spectral energy distribution at $z\approx0.8$ has a colour $(I-K_s) \approx 1.8$, whereas the same galaxy at $z\approx 1.4$ shows $(I-K_s)\approx 3.1$, and reaches $(I-K_s) \geq 4.5$ at redshift $z=2$. This means that, even if the optical detection image is, as is the case in ALHAMBRA, deeper than the corresponding $K_s$ band, at least some of the incompleteness produced by the selection effects can be avoided by using the $K_s$ band to provide the detection image.

In this work we present a new $K_s$-band selected catalogue of galaxies in the ALHAMBRA survey that has been compiled in order to partially overcome the selection bias described above. With this catalogue we will be able to extend some of the works that have already been performed with the ALHAMBRA data to higher redshifts $z>1$, namely: calculation of the general and type-segregated correlation functions \citep{2014MNRAS.441.1783A,2016ApJ...818..174H}, search for groups and clusters \citep{2015MNRAS.452..549A}, analysis of the  clustering signal encoded in the cosmic variance \citep{2015A&A...582A..16L}, stellar populations of galaxies \citep{2015A&A...582A..14D}, detection of high-redshift galaxies \citep{2015A&A...576A..25V}, and possibly also the morphological analysis of some of the brightest targets \citep{2013MNRAS.435.3444P}. We will also use this catalogue to produce large, well-defined, samples of massive galaxies at intermediate redshifts over the redshift range $1<z<2.5$, as well as Balmer jump selected galaxies at $z>1$ \citep[similarly to what was done in ][]{2016A&A...588A.132T}. 

The organization of the paper is as follows: we briefly introduce the ALHAMBRA survey in \S2, and describe the construction of the catalogue in \S3. Section 4 presents the catalogue and its most basic properties. In \S5 we discuss some of the immediate applications of the catalogue, with particular attention to how its use will be important in order to complete (either in terms of redshift or in terms of galaxy types) some of the analyses that have already been published based on the original ALHAMBRA catalogue. Finally \S6 contains our conclusions. In what follows all magnitudes are given in the AB system \citep{1983ApJ...266..713O}, and we use a cosmology with $H_0=100\ h \mathrm{\ km\ s}^{-1}\ \mathrm{Mpc}^{-1},\ \Omega_\mathrm{M}=0.28,\ \Omega_\Lambda=0.72$ \citep{2015arXiv150201589P}.

%__________________________________________________________________

\section{The dataset}

We present in this section the ALHAMBRA Survey, the dataset we have used for the construction of the catalogue. We introduce both the images and the previously published F814W-based ALHAMBRA galaxy catalogue, which will be used as anchor and comparison for our work in the (wide) sample where they overlap.

\subsection{The ALHAMBRA Survey.}

The Advanced, Large, Homogeneous Area, Medium-Band Redshift Astronomical (ALHAMBRA) Survey\footnote{{\tt http://www.alhambrasurvey.com}} has mapped eight separate fields in the Northern hemisphere sky, down to magnitude $I_{814}\approx 25$, using a purpose-built set of twenty 310-\AA\ wide, top-hat, contiguous and non-overlapping filters that cover the whole visible range from $\sim$3500\AA\ to $\sim$9700\AA, plus the standard near-infrared $JHK_{\rm S}$ filters. The survey is fully described in \cite{moles2008}, and the final catalogue can be found in \cite{2014MNRAS.441.2891M}, hereafter M14. Five of the eight observed fields correspond to well-known survey areas (they overlap, respectively, with the DEEP2, COSMOS, HDF, EGS and ELAIS-N1 fields). We refer our readers to the two papers mentioned above for the most accurate details on the project, and only present a brief overview here.
	
The main driver behind the ALHAMBRA Survey was to create a relatively large, deep, and homogeneous catalogue of galaxies with multi-band photometry and high-quality photometric redshifts, that could be used to analyse the processes of galaxy evolution over approximately 50\% of the history of the Universe. The observations were carried out with the 3.5 m telescope of the Centro Astron\'omico Hispano-Alem\'an (CAHA\footnote{{\tt http://www.caha.es}}) in Calar Alto,  Almer\'ia (Spain), where two different cameras were used: the Large Area Imager for Calar Alto (LAICA\footnote{{\tt http://www.caha.es/CAHA/Instruments/LAICA}}) in the optical and OMEGA2000\footnote{{\tt http://www.caha.es/CAHA/Instruments/O2000}} in the NIR. The images were collected between the years 2005 and 2010 and a grand total of $\sim700$ hours of on-target observing time was compiled, for a total effective survey area of $\sim2.8$ square degrees. The final catalogue presented in M14 includes $\sim438,000$ galaxies with $\left\langle z \right\rangle =0.86$ and rms photometric redshift accuracy $\delta z / (1+z) =0.014$.
	
In order to produce a sample that could be comparable to other surveys, a synthetic detection image was created for every field using the medium-band ALHAMBRA images. This image corresponds very accurately to the one that would be obtained using the HST filter F814W, and we will refer to it over this work as $I_{814}$ for simplicity, even though it does not exactly correspond to the usual Johnson $I$ band. This synthetic image was used for object detection, thus producing object lists and photometric catalogues that are magnitude-limited in the $I_{814}$ band. These catalogues were carefully compared to the ones obtained by \cite{2009ApJ...690.1236I} in the COSMOS field, proving the validity of the approach.	
		
\subsection{ALHAMBRA $K_s$-band images}

In order to add information in the NIR range of the spectral energy distributions, the three broad-band standard $JHK_s$  filters were included in the survey. Having these three filters in the infrared range helps in breaking the well-known degeneracy between the 4000 \AA\ break at low redshift and the Lyman break in more distant galaxies. Furthermore the extra information provided significantly increases the scientific value of the data, particularly for elliptical galaxies, strongly reddened AGN, or moderate-redshift starburst galaxies.
	
The NIR images also provide a set of sources that are not included in the ALHAMBRA main catalogue because of their very red colours, thus in this work we present a new $K_s$-band selected catalogue. From the very early phases of the ALHAMBRA NIR data reduction \citep{2009ApJ...696.1554C} we noticed that this subset of our data was interesting by itself. Visual inspection and comparison of the $K_s$-band data with the images in the visible range showed that the former, although obviously shallower than the average of the latter, contained a sizeable sample of objects whose red $(I_{814}-K_s)$ colours made them more noticeable in the NIR images. 
    
\begin{figure}
\centering
\includegraphics[width=0.45\textwidth]{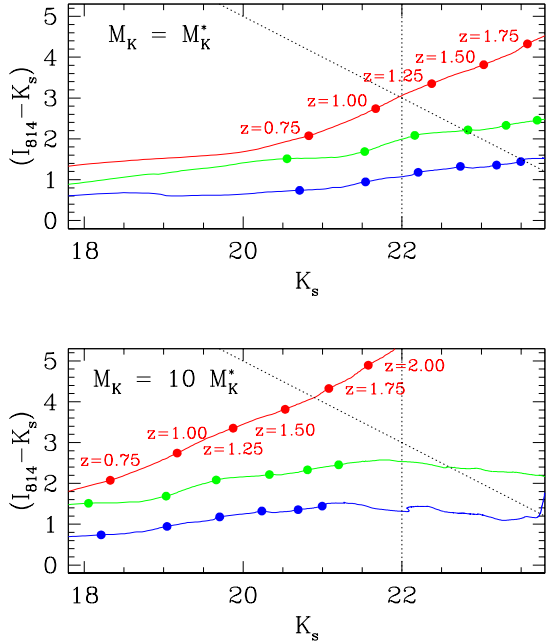}
\caption{ ($I_{814}-K_s$) vs. $K_s$ colour-magnitude diagram. In each plot the (red, green, blue) (also top, medium, bottom) track corresponds to an (elliptical, spiral, starburst) galaxy template. Some redshift values are marked on the elliptical template track as a reference.}\label{colcoltheo}
\end{figure}

 Figure \ref{colcoltheo} shows a theoretical colour-magnitude diagram, with the redshift tracks of different ${M_K^*}$ (top) and $10 M_K^*$ (bottom) galaxy templates plotted on a ($I_{814}-K_s$) vs. $K_s$ plane. It is designed to show the expected reach of the regular ALHAMBRA catalogue and that of a $K_s$-band selected one, in order to offer the reader a visual intuition of the main objective of this work. We have allowed for an evolving value of $M_K^* = -22.2 - 0.5(1+z)$, which is an approximation derived from the luminosity function analyses by \citet{2007A&A...476..137A} and \citet{2006MNRAS.367..349S}, and references therein. The limit magnitude values plotted on the diagram correspond to $K_s=22.0$, $I_{814}=25$. There is an obvious gain in depth for intrinsically red objects when the near-infrared images are taken as reference (vertical dotted line) compared to a $I_{814}$-selected sample (diagonal line), particularly in the case of luminous red objects at redshift $z>1$. For example , for the nominal values in the plot (which is only an approximation) the reach of the survey for a 10$M^*$ early-type galaxy would change from $z \approx 1.65$ ($I_{814}<25$) to $z \approx 2.20$ ($K_s<22.0$).
    
%%%%%%%%%
%%%%%%%%%

\begin{figure}
\centering
\includegraphics[width=0.45\textwidth]{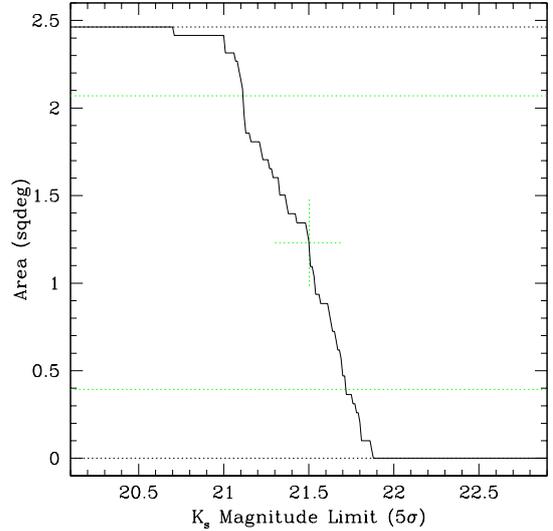}
\caption{Cumulative area covered by the ALHAMBRA $K_s$-band images as a function of the magnitude limit reached in each one. This calculation uses the nominal 5$\sigma$ limit in each pointing, and has been corrected using the image masks described in Section \ref{masks}. The horizontal dotted lines mark the area values corresponding to 0, 16, 84 and 100\% of the total survey area.} \label{figklim}
\end{figure}

\label{maglims} 
The conditions under which the NIR observations of the different ALHAMBRA fields and pointings were observed were varying, which leads to a clear and significant non-uniformity in the magnitude detection limits for each of them. The median limit\footnote{Magnitude limits quoted here are nominal 5$\sigma$ limits measured in circular, 3-arcsec diameter apertures.} of the $K_s$-band images is AB=21.5, with 68\% of the images having a $5\sigma$ limiting magnitude value between 21.1 and 21.7, as seen in Figure \ref{figklim}.

%%%%%%%%%
%%%%%%%%%

\subsection{Data reduction}
\label{sec:datared}
The ALHAMBRA images, as mentioned, have been taken in eight different fields\footnote{Only seven have been completely observed and reduced, with ALHAMBRA-1 being unfinished at this stage.}. Given the particular structure of the LAICA focal plane, consisting of four detectors, each one covering approximately 15x15 arcmin squared, whose centers are situated at the corners of a (virtual) 30x30 arcmin square, one pointing includes four such images. Two neighbouring LAICA pointings produce two horizontal strips in the sky, each of them measuring 60x15 arcmin and separated by 15 arcmin in the vertical direction (see Section \ref{masks} for details). This is the shape of each of the ALHAMBRA fields\footnote{Except for fields ALHAMBRA-4 and ALHAMBRA-5 for which we have only covered in full one LAICA ponting.}. The basic unit in the ALHAMBRA reduction and analysis is a LAICA CCD, which we identify, for example, as F04P01C01 for CCD\#1 in the first pointing of the ALHAMBRA-4 field. A full illustration can be seen in Appendix A of M14.

The NIR images provided by the camera OMEGA2000 cover the same area of a single LAICA CCD, and had an original pixel scale of $\sim 0.45$ arcsec pixel$^{-1}$. In order to supply an homogeneous dataset, the NIR images were re-sampled to the LAICA pixel scale ($\sim 0.225$  arcsec pixel$^{-1}$), which represented an interpolation over a $2\times2$ grid per pixel. The individual images were dark-corrected, flat-fielded, and sky-subtracted, and individual masks were created to account for bad pixels, cosmic rays, linear patterns, blemishes, and ghost images coming from bright stars. The SWarp Software \citep{2002ASPC..281..228B} was used to combine the processed images correcting geometrical distortions in the individual images, using the astrometric calibrations stored in their World Coordinate System (WCS) headers. Once the images were combined, a preliminary source catalogue for each pointing was created and cross-matched with the Two Micron All Sky Survey  \citep[2MASS,][]{2003yCat.2246....0C} in order to select common objects with high S/N, to be used for calibration purposes. In this work, we present the results obtained using the re-sampled, corrected, combined images. The full, detailed description of the reduction process is given in \cite{2009ApJ...696.1554C}. 

For each combined and fully calibrated NIR image we have used the same method presented in \cite{2014MNRAS.441.1783A} to calculate an associated pixel mask that accounts for possible remnant cosmetic problems, not homogeneously covered image borders, and saturated stars. After masking, the total area covered by our catalogue amounts to 2.463 square degrees.

Flux calibration of the 20 medium-band optical filter images was achieved using relatively bright stars in each of the CCDs as secondary standards, and anchoring them to their Sloan Digital Sky Survey photometry \citep{2000AJ....120.1579Y}. Flux calibration of the $JHK_s$ images was based directly on 2MASS \citep{2003yCat.2246....0C}. Full details of the process and the accuracy reached in every case can be found in \citet{2009ApJ...696.1554C}, \citet{2010AJ....139.1242A}, and Cristobal-Hornillos et al. (in preparation).

%__________________________________________________________________

\section{Construction of the catalogue}

We present in this section the process leading to the construction of the catalogue, including image detection, photometry in the reference $K_s$ band, and in the rest of the ALHAMBRA filters, angular selection mask, calculation of the completeness functions, star-galaxy separation, and photometric redshift estimation. We leave for the next section the discussion of the basic properties of the catalogue and the checks we have performed in comparison with other available data to test the quality and reach of our results.

We have tried to keep the process used to generate the $K_s$-band catalogue as close as possible to the one that was performed by M14 over the $I_{814}$ images both to improve our ability to compare the results and also to keep some degree of consistency between them. Thus, we will often refer to that work for details.  

\subsection{Source detection and photometry}

Source detection was performed using SExtractor \citep{Bertin1996} over each of the $K_s$-band images. As it is usual in this kind of work, in order to optimize the number of real sources we performed detection both on the original images and their negatives using different sets of parameters, exploring parameter space to maximize the number of real detections while securing the least possible spurious ones. We finally opted for a minimum area of 5 connected pixels with signal greater than 1.2 times that of the background noise, after filtering with a 5-pixel FWHM Gaussian kernel\footnote{That is, \texttt{DETECT\_THRESHOLD} $=1.2$ and \texttt{DETECT\_MINAREA} $=5$}. At first order this would imply a minimum S/N $\gtrsim 10$ for the detected sources. As already mentioned in Section \ref{maglims} the median $5\sigma$ limiting magnitude of our images is $K_s\simeq 21.5$. A more detailed and realistic analysis of the photometric depth is explained below.
     
Photometry was then carried out over the 20+3 ALHAMBRA images plus the synthetic F814W one using SExtractor in dual image mode, using the $K_s$-band images for detection in every case. We introduced as input to SExtractor the values of the zero points for each image, as calibrated during the reduction process.  We changed the values of the parameters \texttt{DETECT\_THRESHOLD} and \texttt{DETECT\_MINAREA} in the photometry mode in order to define the photometric apertures in the same way as was done in the case of the ALHAMBRA optical catalogue We remark that in ALHAMBRA, as is usually done in multi-band surveys, one image is used for the detection of the objects and the definition of the isophotal apertures (a synthetic $F814W$ in ALHAMBRA, the $K_s$ band in our case), which are then used to measure the isophotal flux of each object in all the other bands. Obviously the individual apertures themselves will be different if both catalogues were compared, as (i) the brightest part of each galaxy, that defines the aperture, will be intrinsically different in the $K_s$ and the F814W bands, and (ii) the seeing in the NIR images is generally and significantly better than that in the optical ones\footnote{As shown in M14, the median seeing is $\sim0.9''$ for the NIR images, $\sim1.1''$ for the visible images, and $\sim1.0''$ for the F814W synthetic images---which are generated selecting preferentially those with good seeing within the adequate wavelength range.}. One of the most important checks that we will perform on the final catalogue will be devoted to check the compatibility between the general-purpose ALHAMBRA catalogue photometry and our own in those objects they have in common.

\subsection{Photometric errors}
\label{photerr}

A proper estimation of the photometric errors represents an important task for photometric redshift estimation, since the techniques used to compute them rely heavily on the photometric uncertainties. When SExtractor estimates the photometric uncertainties, it assumes that the noise properties are characterized by a Poisson distribution. This is correct only if there are no correlations between pixels. 
    
The reduction and re-sampling processes executed on the NIR images cause significant correlations between pixels, and the assumption of a standard Poisson estimation of the background noise leads to a significant underestimation of the real photometric errors. This underestimation is aggravated in the case of faint sources \citep{2009ApJ...696.1554C}. The final flux error  for each source was calculated as:
\begin{equation} 
\sigma^2_F=\Big[\sigma_0 \cdot K \cdot \sqrt{A}\cdot \Big(a+b\sqrt{A}\Big)\Big]^2+\Bigg(\frac{K^2 \cdot F}{G}\Bigg)+\Sigma_{\rm Phot Calib}^2,
\label{fitabpar2}
\end{equation}
where the first term is the background error estimated following the method used by \cite{2003AJ....125.1107L}, with $K$ being the value of the weight map\footnote{The weight map measures the relative exposure time per pixel within the same pointing for a given filter.} in the region where the source is measured, $\sigma_0$ the background RMS, and $A$ the area of the aperture in each case, as given by the \texttt{ISOAREA\_IMAGE} SExtractor output value. The term that includes the $a$ parameter encloses the errors due to correlation between neighbouring pixels, while the $b$ parameter term includes the large-scale correlated variations in the background. Both $a$ and $b$ were obtained as a result of the fitting process of measured the standard deviation, corresponding to the fluxes measured in each ALHAMBRA background image in different size boxes, as in \cite{2003AJ....125.1107L}. The second term was added in order to estimate the shot noise error related to the source flux F and the gain G. The third term is the error due to the calibration uncertainty $\Sigma_{\rm PhotCalib}$.

Finally, the magnitude uncertainties were calculated applying the equation:
\begin{equation} 
\mathrm{\sigma_M = 1.0857 \cdot \frac{\sigma_F}{F}}.
\label{magerr}
\end{equation}

\subsection{Angular selection mask}
\label{masks}

In order to take into account possible position-dependent selection effects, we built an updated version (v2) of the ALHAMBRA survey angular selection mask presented in \citet{2014MNRAS.441.1783A}. These masks were built to define the sky area which has been reliably observed, excluding regions with potential problems. The latter include regions with low exposure time next to the borders of the CCDs, regions next to bright stars or saturated objects and regions where obvious defects in the images are found \citep[for details, see ][]{2014MNRAS.441.1783A}.

\begin{figure*}
\centering
\includegraphics[width=\columnwidth]{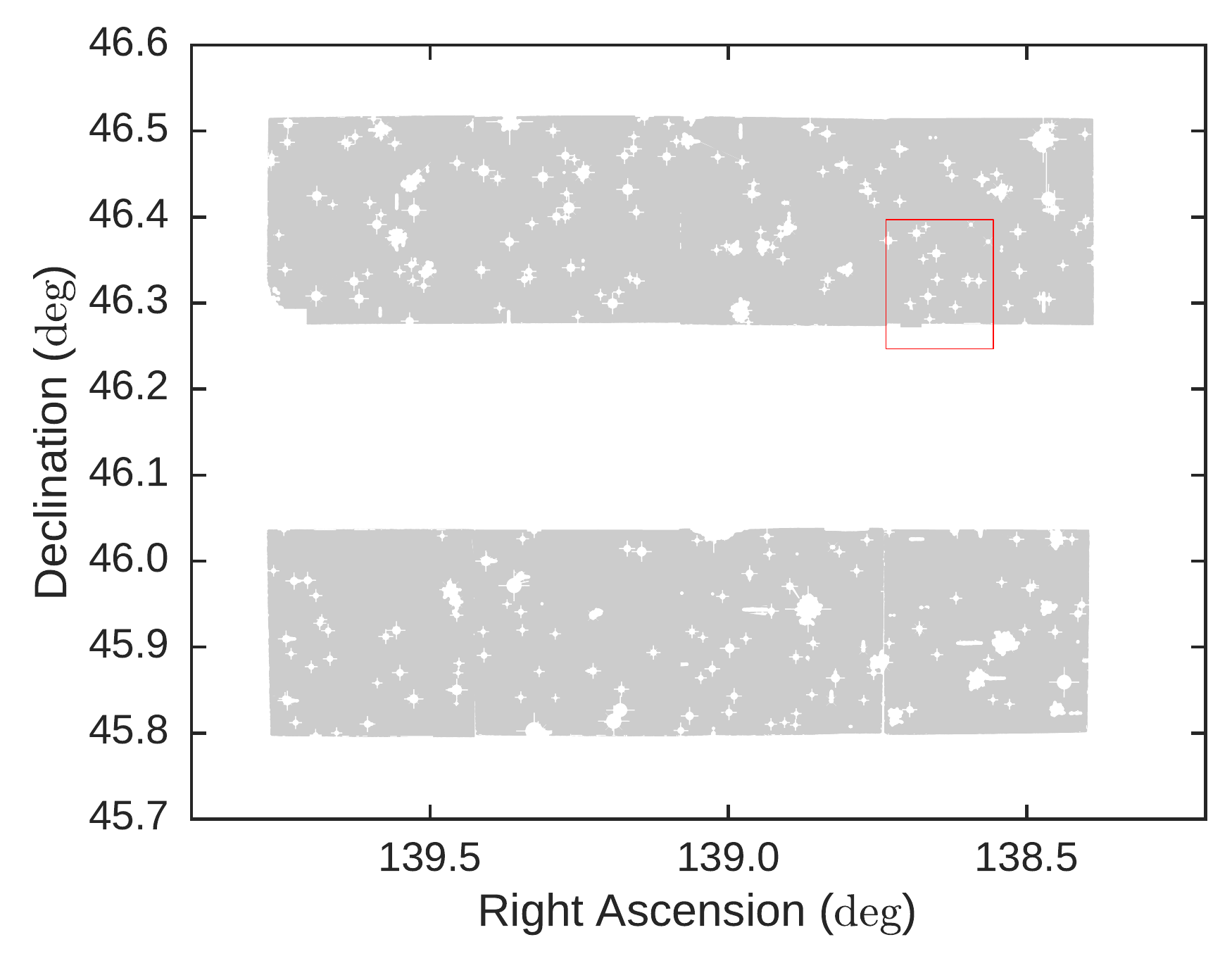}
\includegraphics[width=\columnwidth]{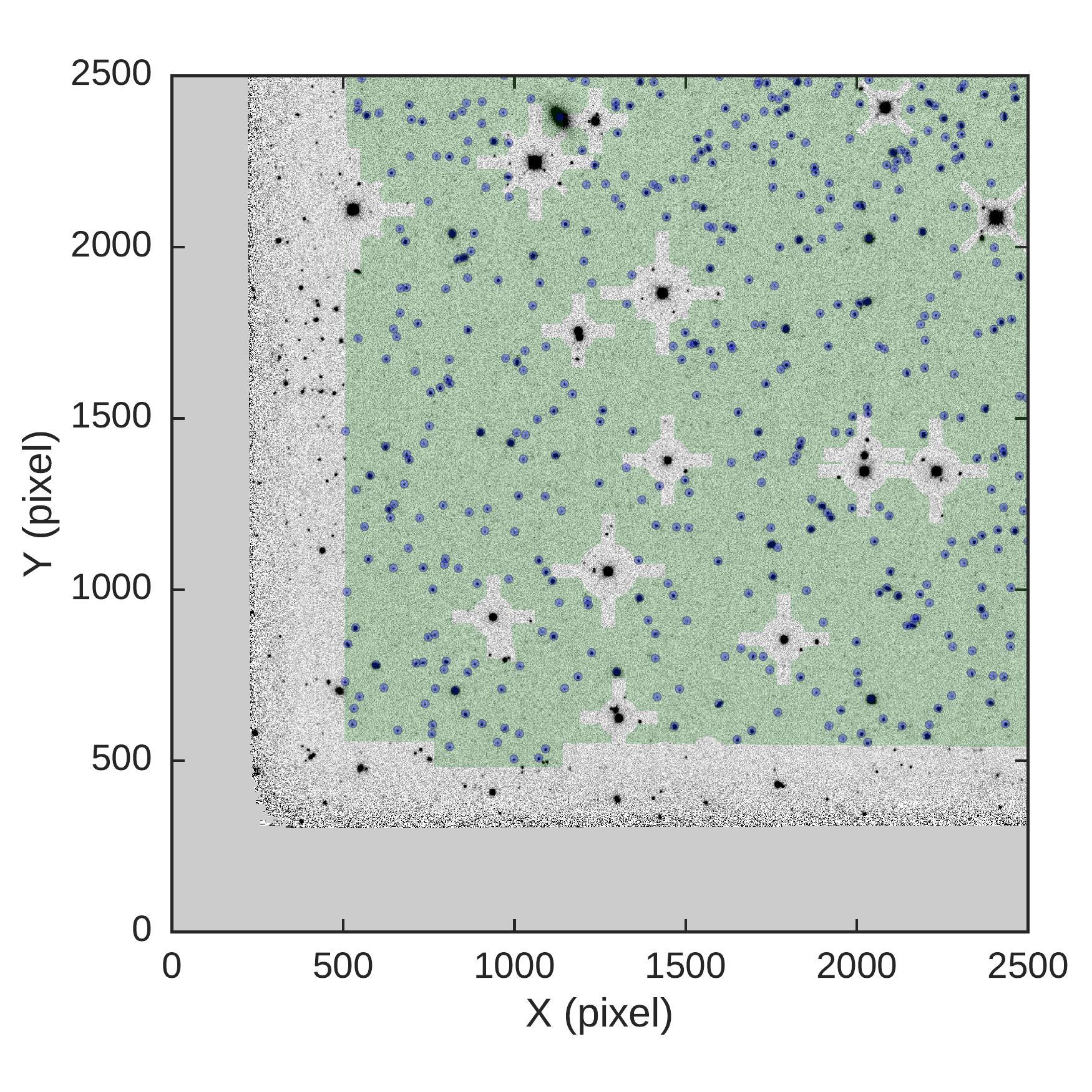}
\caption{Illustration of the ALHAMBRA survey angular mask (v2) for the ALHAMBRA-3 field. 
        Left: angular mask for the complete field. The shaded area corresponds to the region included in the selection. This shows the peculiar ALHAMBRA field geometry described in Sect.~\ref{sec:datared} and in M14. The red rectangle marks the area shown in the right image.
		Right: Detail of one typical ALHAMBRA $K_s$ image, showing the corresponding angular selection mask (shaded in green). The blue points correspond to detected objects included in our catalogue. This image corresponds to an area of $\sim 8 \times 8$ arcmin$^2$. 
        We see how regions near the border of the CCD image are excluded from the mask. 
        We also exclude a cross-shaped region around each saturated object, with the vertical/horizontal crosses corresponding to diffraction spikes in the optical F814W image, and those at $45^{\circ}$ corresponding to spikes in the $K_s$ image.} \label{fig:mask}
\end{figure*}

In this new version we built two different masks following this approach, one based on the synthetic F814W images, and the other based on the $K_s$ images. 
The optical-based angular mask is very similar to the one presented in \citet{2014MNRAS.441.1783A}, with two small differences. First, we have used an updated version of the \emph{flag} images that describe the regions with appropriate effective exposure times. These now include some small areas (mainly in field ALHAMBRA-2) that were previously incorrectly excluded. Second, we now mask out regions around bright and saturated stars using a shape that properly matches the diffraction spikes (see Figure \ref{fig:mask}). The NIR-based angular mask was created following the same approach, but based on the map of effective exposure times and saturated objects in the $K_s$ images. We take into account the fact that, due to the different disposition of the LAICA and OMEGA2000 cameras, the orientation of the diffraction spikes is rotated $45^{\circ}$ between the optical and NIR images.

We combined the optical- and NIR-based masks into a final mask that therefore describes the sky region that has been reliably observed both in the optical and in the NIR. From this final mask we excluded some small regions to avoid overlap between neighbouring CCDs. Figure \ref{fig:mask} is an illustration of the resulting ALHAMBRA survey mask (v2) for one of the fields (ALHAMBRA-3). The total effective area of the survey according to this angular selection mask is $A_{\rm eff} = 2.463 \deg^2$, and the effective areas for each of the fields are listed in Table~\ref{table1}. The small increase ($\sim 3\%$) in area with respect to v1 of the masks \citep{2014MNRAS.441.1783A} is due to the aforementioned differences in the optical-based masks.

Even after the masks were applied over the images we observed small residual, periodic electronic ghost patterns over detector rows and columns around very bright, saturated stars in the $K_s$-band images. We individually checked and removed a total of 59 sources in these problematic areas from the catalogue.

The angular masks were generated using the \textsc{Mangle} software \citep{swa08a}, and we will make them publicly available (in \textsc{Mangle}'s \emph{Polygon} format) together with the data catalogue.
We list in the data catalogue all objects detected in the full $K_s$-band images, and column \texttt{MASK\_SELECTION} in the catalogue indicates whether a given object is inside the angular selection mask or not.
All the analyses performed in this paper consider only the objects inside the mask (with \texttt{MASK\_SELECTION = 1} in the catalogue), and this is the approach we recommend for any further statistical analysis based on this catalogue.

 %%%%%%%%%
%%%%%%%%% 

\subsection{$K_s$-band completeness}
\label{completeness} 

A key ingredient for any analysis to be performed with the catalogue is the measurement of its completeness. As was described above, our survey includes 48 independent images, distributed over seven different fields. Each one of them was observed and analysed using the same parameters, exposure times and instruments. However, the observing conditions in each case were very different: the period of time over which the observations took place covered several years during which the instruments passed successive cycles. Obviously the atmospheric conditions were also widely different between the observing runs.

Therefore the limiting magnitudes that define the depth of our catalogue vary widely from one field to another, and also within the different CCDs in the same field. In order to minimize this effect we need to estimate a completeness function that will allow us to compute the corrections at the faint end of the galaxy number counts. 

We have already mentioned that the ALHAMBRA pointings were chosen to overlap with well-known fields. In particular, an area of $\sim$ 0.21 $\mathrm{deg^2}$ of the ALHAMBRA-4 field overlaps with the UltraVISTA COSMOS field \citep{2012A&A...544A.156M}. Since the magnitude limit in the UltraVISTA $K_s$-band selected catalogue for this field is $AB \sim 24$ \citep{2013ApJS..206....8M}, and the ALHAMBRA magnitude limit is $AB \sim 22$, we can estimate our $K_s$-band completeness function using the UltraVISTA COSMOS data as reference.
 
\begin{figure}
\centering
\includegraphics[width=\columnwidth]{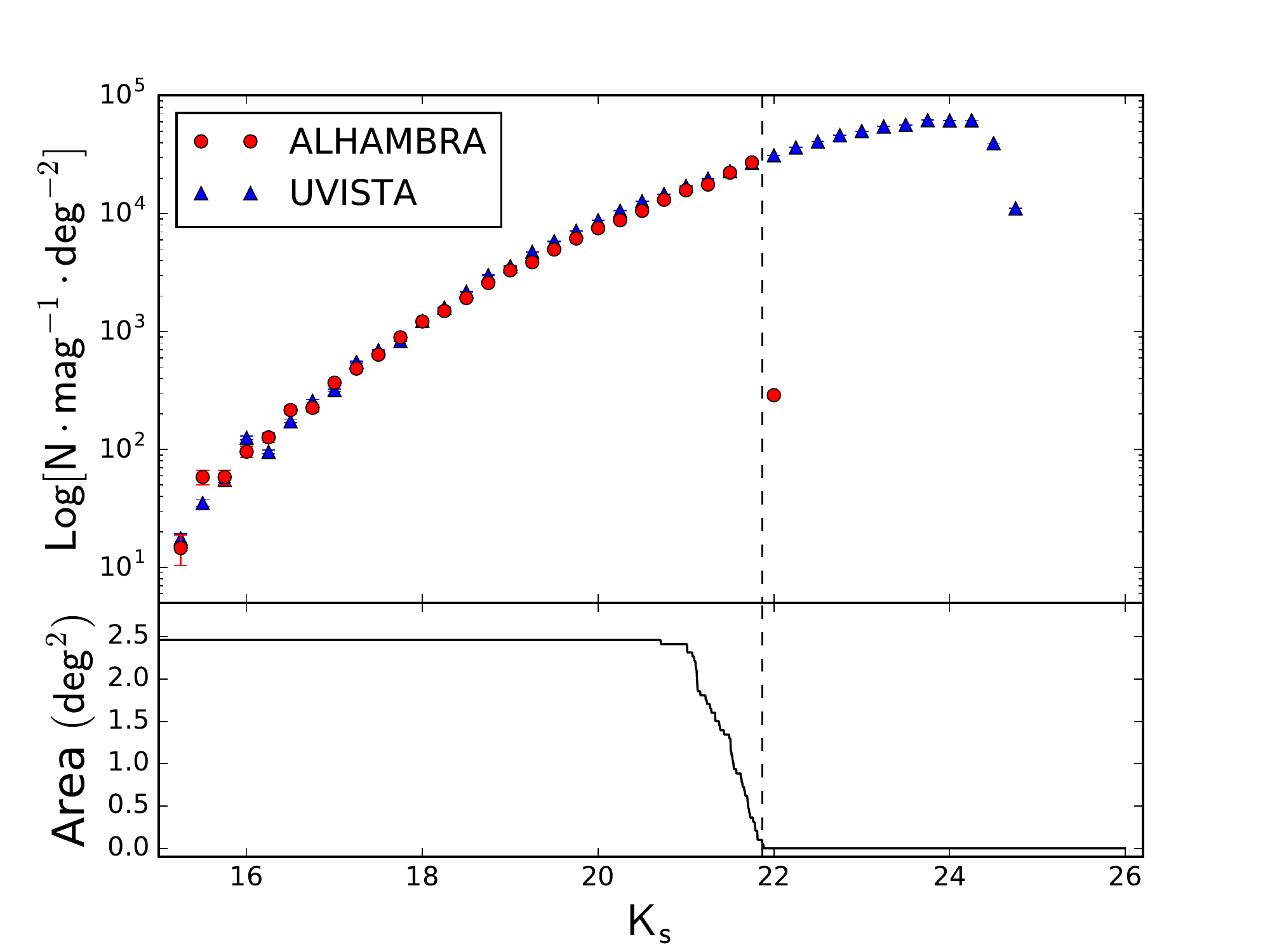}
\caption{Corrected galaxy number counts in the $K_s$-band ALHAMBRA catalogue. The vertical dashed line marks the magnitude at which the completeness falls to 60\% for the deepest images. The lower panel shows the available survey area at each magnitude, using as limit for each CCD the value $(m_{5\sigma}+m_c)$ of the half-completeness point in the Fermi function as described in Appendix \ref{appcompleteness}.}\label{NumberCounts}
\end{figure}

The complete procedure is described in detail in Appendix \ref{appcompleteness}, and we only show here the results obtained when the (CCD-dependent) completeness correction is applied to each pointing and the final result is compiled. Figure \ref{NumberCounts} shows the result of such procedure with the total counts in our catalogue compared to those in the deeper UltraVISTA sample.
  
\subsection{Star-galaxy separation}
\label{stargal}

SExtractor outputs for each object in the catalogue a value for the \texttt{CLASS\_STAR} parameter. This parameter estimates the stellarity of each source attending to morphological criteria. However, given the average seeing of the ALHAMBRA images, this value is not trustworthy for most of the objects in the catalogue. Moreover, even if the average seeing in our images were much better, we may still face cases of compact galaxies which could be morphologically misidentified as stars. We must thus apply an additional classifying method to improve our star-galaxy separation. 
	
\begin{figure*}
\centering
\includegraphics[width=0.45\textwidth]{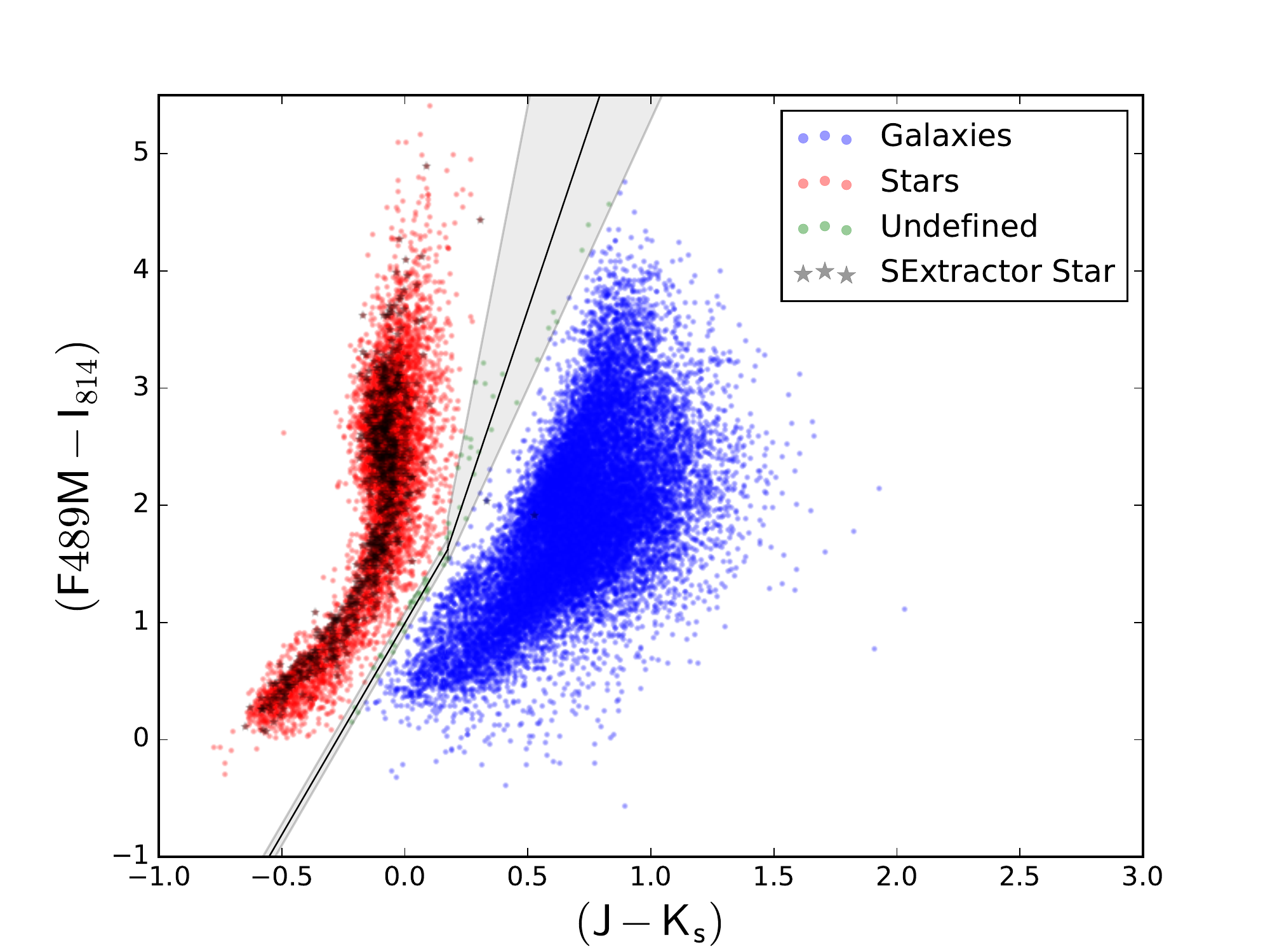}
\includegraphics[width=0.45\textwidth]{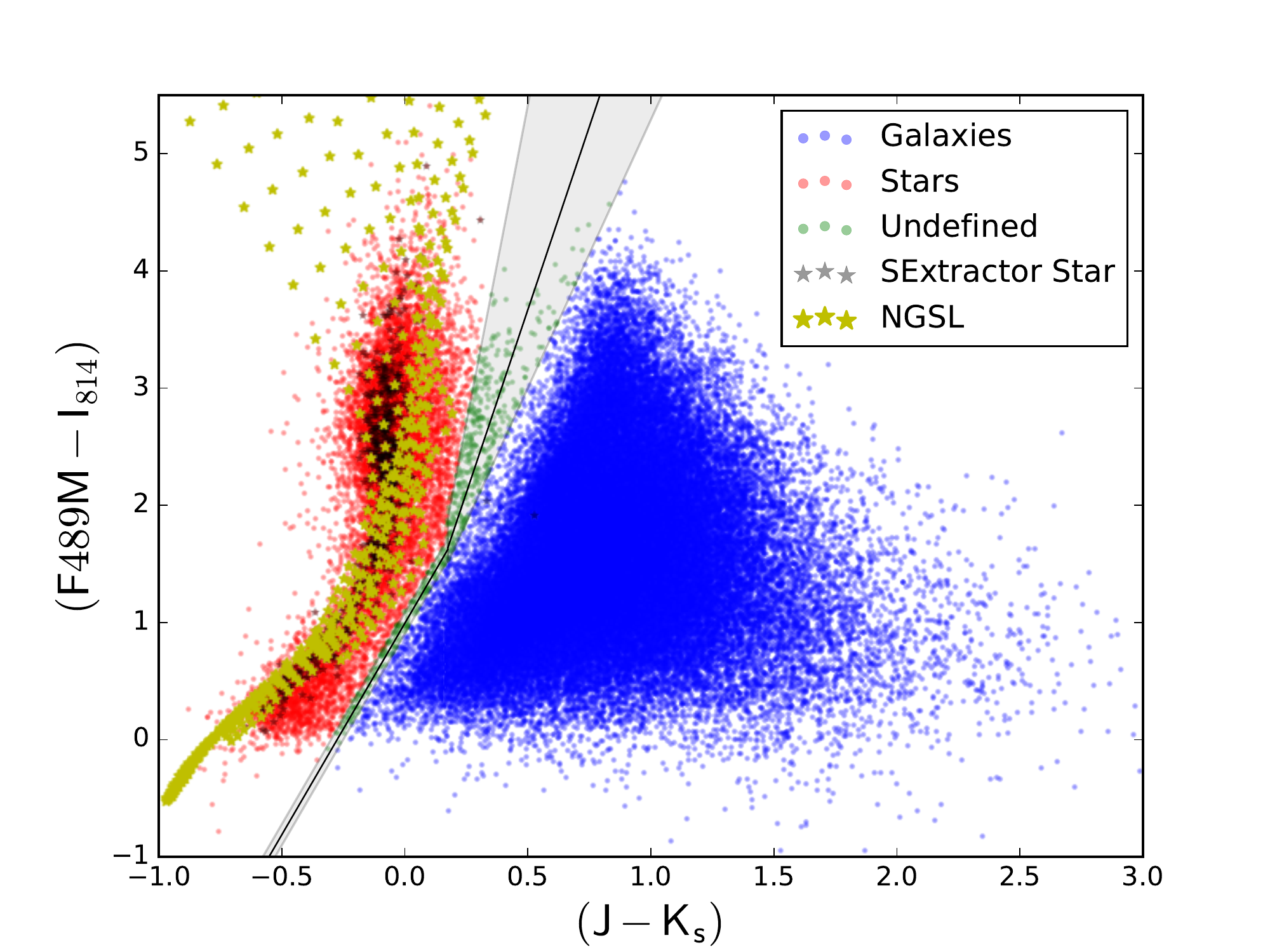}
\caption{Colour-colour diagram used to perform the photometry-based star-galaxy separation. In the left panel we restrict the plot to $K_s<18$ in order to show clearly the different loci occupied by stars (red) and galaxies (blue), as well as the objects classified as stellar by SExtractor (black). The right panel shows the same diagram for the whole catalogue. The grey area encloses the objects that are not securely identified either as galaxies or stars, and the amber markers correspond to the positions of stars included in the NGSL.} \label{galstar}
\end{figure*}

As described in \cite{1997ApJ...476...12H}, a colour-colour diagram combining near-infrared and visible colours can provide a simple albeit accurate criterion to discriminate between stars and galaxies. We will use the colour (F489M-$I_{814}$) in the optical range and the $(J-K_s)$ colour in the NIR. Figure \ref{galstar} (left) shows such a colour-colour plot where we have applied a magnitude selection limit $K_s<18$ in order to avoid any dispersion due to large photometric uncertainties and see the stellar locus as a well-defined area. 

The line that separates the loci corresponding to galaxies and stars is marked on the plot and given by:
%%%%%%%%%
\begin{multline}
\mathrm{F489M}-I_{814}=3.61*[(J-K_s)+0.275]\\ \hfill [(J-K_s)<0.17]
\label{recta1}
\end{multline}
%%%%%%%%%
\begin{multline} 
\mathrm{F489M}-I_{814}=6.25*[(J-K_s)+0.087]\\ \hfill [(J-K_s) \geq 0.17]
\label{recta2}
\end{multline}
%%%%%%%%%	
Black stars mark the objects that SExtractor classifies as stellar (\texttt{CLASS\_STAR} $>0.95$) in the range where such classification is accurate ($K_s<18$). In the right panel on Figure \ref{galstar} we show the same colour-colour diagram applied to our whole sample. As a further check, we have also included amber markers at  the positions where stars in the Next Generation Spectral Library \citep[HST/STIS NGSL,][]{2004AAS...205.9406G} would fall. 

We have included in our catalogue a column called \texttt{ COLOR\_CLASS\_STAR}, which takes the value 0 for objects classified as galaxies using this diagram and 1 for those classified as stars. We have also defined in the colour-colour diagram an area where classification is not clean, within which we have assigned the value 0.5 to all objects---they are  marked in green and they fill the grey area in Figure \ref{galstar} (right). In addition, we have compared the stellarity of the sources in this work and M14, finding that less than 0.1\% of the common sources present inconsistencies.

%%%%%%%%%	
%%%%%%%%%

\subsection{Photometric Redshifts}
\label{photoz}

Photometric redshifts for all galaxies in the catalogue have been calculated using the Bayesian Photometric Redshift code \citep[BPZ2.0,][Ben\'itez in preparation]{2000ApJ...536..571B}. We refer the reader to M14 and \cite{2000ApJ...536..571B} for further details on BPZ and its application, and list here only some of the most basic properties of the code.

A total of 11 different galaxy spectral energy distribution (SED) templates were used: five from elliptical galaxies, two for spiral galaxies and four for star-burst galaxies. The SED types are numbered following the same sequence from $T_B=1$ to $T_B=11$. The spectral fitting includes emission lines and dust extinction within the templates themselves, and not as separate parameters. Linear interpolation between the types was included in order to improve the coverage of SED-space and make it denser. BPZ calculates the likelihood of the observed photometry for all the combinations of redshift and SED type in the given parameter space, and combines it within a Bayesian formalism with priors calculated as distributions of the density of the different spectral types as a function of redshift and magnitude. The output of the code includes both best-fitting solutions, one coming from the likelihood analysis alone and the second one including the prior information. It also outputs the full probability distribution function PDF$(T_B,z)$ which should be used preferentially for the ensuing analyses.

BPZ calculates an extra parameter which will be very important for us: the {\it Odds} parameter, which corresponds to the integration of the PDF within a narrow redshift range around the best-fitting solution. High values of the {\it Odds} parameter mark objects whose redshift is very well determined, with a narrow, single peak in the probability distribution. Low values of {\it Odds} signal either objects that due to poor-quality photometry or to a lack of an adequate SED in our library of templates suffer a poor fitting; or objects that inhabit an area of colour space which has an intrinsic degeneracy between two different redshifts\footnote{This should in fact be a minimal problem for ALHAMBRA because of the 23 photometric bands that are used, but can be more serious in our case because for very red objects we are sometimes left with detections only in a few of the reddest filters.}. 

Our catalogue includes as output from BPZ 2.0: the photometric redshift Bayesian estimate ${z_b}$, the associated SED best-fitting type $T_B$, the {\it Odds} parameter, the maximum-likelihood estimates of redshift and SED type, and some derived measurements like absolute magnitudes and an estimate of the stellar mass\footnote{The stellar mass is a rough estimate, derived from the flux normalisation and SED type.}.

\section{Catalogue Properties}

We have performed a series of tests on our catalogue to ensure its accuracy and validity. In this section we present some of them, as well as a brief insight of some of its forthcoming applications.

\subsection{Catalogue counts}
The complete catalogue includes photometry for 94,182 sources. They are distributed in the ALHAMBRA fields as shown in Table \ref{table1}. The derived density is $\sim 38,000$ sources per square degree.

\begin{table*}
\centering
\caption{\label{table1} ALHAMBRA $K_s$-band catalogue counts.}
\begin{tabular}{lcccccccc} 
\hline
	Field name			&RA			& DEC 		& Area & Area  	&Sources& Sources& Sources/deg$^2$  &Magnitude Limit\\ 
    					&(J2000) 	& (J2000) 	&(full) deg$^2$ & (masked) deg$^2$ & (full)  &(masked) & (masked) & $5 \sigma $ \\ 
\hline
	ALHAMBRA-2/DEEP2	& 02 28 32.0& +00 47 00 & 0.441 & 0.402 & 19989 & 16546 & $4.12 \cdot 10^4$ & 21.53\\	
	ALHAMBRA-3/SDSS		& 09 16 20.0& +46 02 20	& 0.500 & 0.415 & 19489 & 16654 & $4.01 \cdot 10^4$ & 21.70\\	
	ALHAMBRA-4/COSMOS 	& 10 00 28.6& +02 12 21	& 0.250 & 0.209 & 11154 &  9587 & $4.59 \cdot 10^4$ & 21.68\\
	ALHAMBRA-5/HDF-N 	& 12 35 00.0& +61 57 00	& 0.250 & 0.218 & 9528  &  8549 & $3.92 \cdot 10^4$ & 21.65\\	
	ALHAMBRA-6/GROTH	& 14 16 38.0& +52 25 05	& 0.500 & 0.415 & 17051 & 14565 & $3.51 \cdot 10^4$ & 21.32\\
	ALHAMBRA-7/ELAIS-N1	& 16 12 10.0& +54 30 00	& 0.500 & 0.414 & 18045 & 15262 & $3.69 \cdot 10^4$ & 21.11\\	
	ALHAMBRA-8/SDSS 	& 23 45 50.0& +15 34 50	& 0.500 & 0.390 & 16116 & 13019 & $3.34 \cdot 10^4$ & 21.19\\  \\							&			& TOTAL		& 2.941 & 2.463 & 111372& 94182 &$3.82 \cdot 10^4$ \\
\hline
\end{tabular}
\end{table*}

\begin{figure*}
\centering
\includegraphics[width=0.45\textwidth]{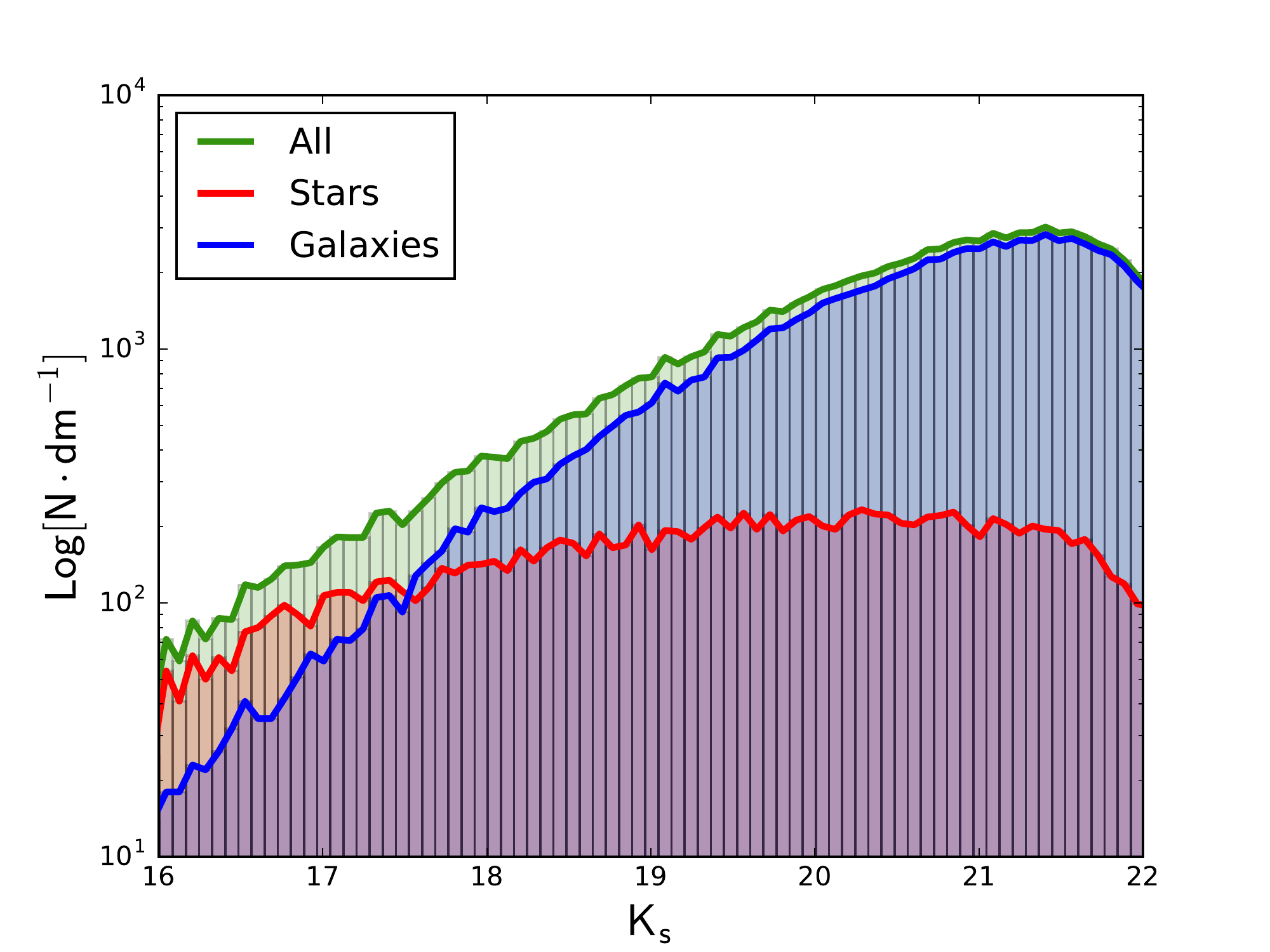}
\includegraphics[width=0.45\textwidth]{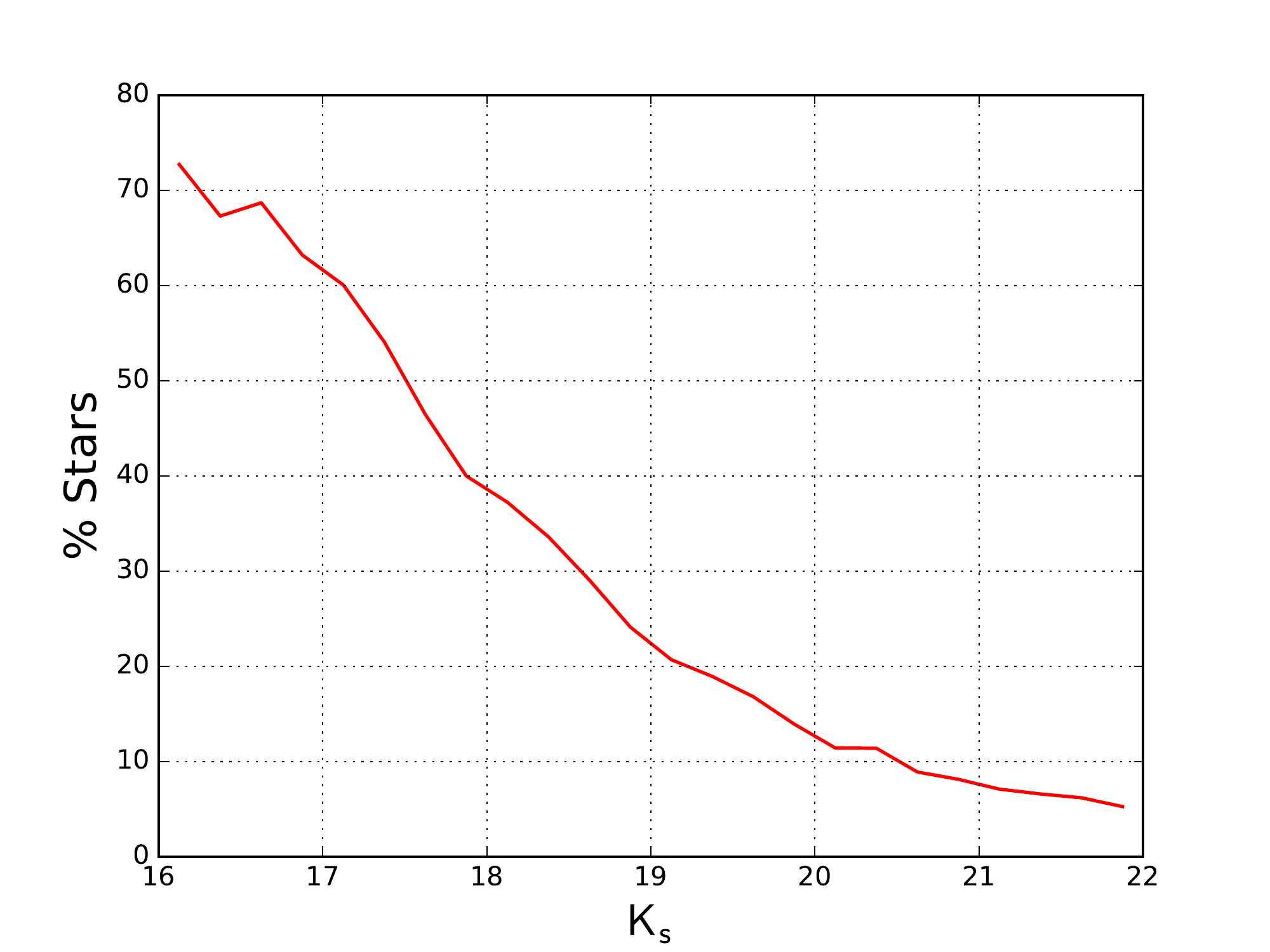}
\caption{Left panel: Raw number of detected sources on the $K_s$ image. The green line corresponds to the full sample, whereas the red and blue colours correspond to the star and galaxy counts, respectively. Right panel: Percentage of stars in the total sample as a function of magnitude.} \label{StarFrac}
\end{figure*}

In the right-hand panel of Figure \ref{StarFrac} we show the histogram of the raw $K_s$-band magnitude counts. We have separated stars and galaxies using the star-galaxy classifier described in Section \ref{stargal}. As expected, stars are dominant until $K_s \approx 17.5$. From this magnitude on, the galaxy fraction increasingly dominates the counts. In the left panel of Figure \ref{StarFrac} we see how the fraction of stars falls, representing less than 10\% of our counts from $K_s \approx 20.5$.
       
%%%%%%%%%	
%%%%%%%%%

The left panel of this latter plot can be compared to Figure \ref{NumberCounts}, where we  showed the galaxy number density vs magnitude plot, once the completeness correction calculated in Section \ref{completeness} is applied. In that case we can extend the range over which the counts are accurate out to $K_s \approx$ 22. Comparison with previous works shows that the counts are consistent, and allows us to perform the following tests. 

\subsection{Colour-magnitude diagram}

Our main motivation to provide a $K_s$-band selected sample is to cover the area of the colour-magnitude diagram where sources with high $(I_{814}-K_s)$ colour reside. These objects are detected in the $K_s$-band catalogue, but many of them have barely any signal in the F814W images. In fact, as expected when a deeper image is used to detect objects and perform photometry in a second band, many objects that went undetected in the original catalogue (because their detected flux did not reach the minimum necessary to fulfill the detection criteria) do have positive flux once the apertures are defined with a second deeper/redder band. 

\begin{figure*}
\centering
\includegraphics[width=0.45\textwidth]{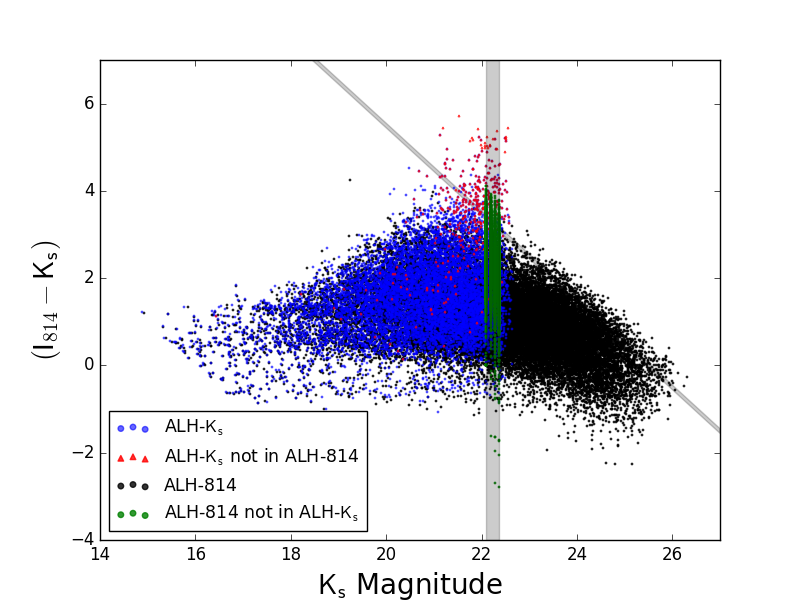}
\includegraphics[width=0.45\textwidth]{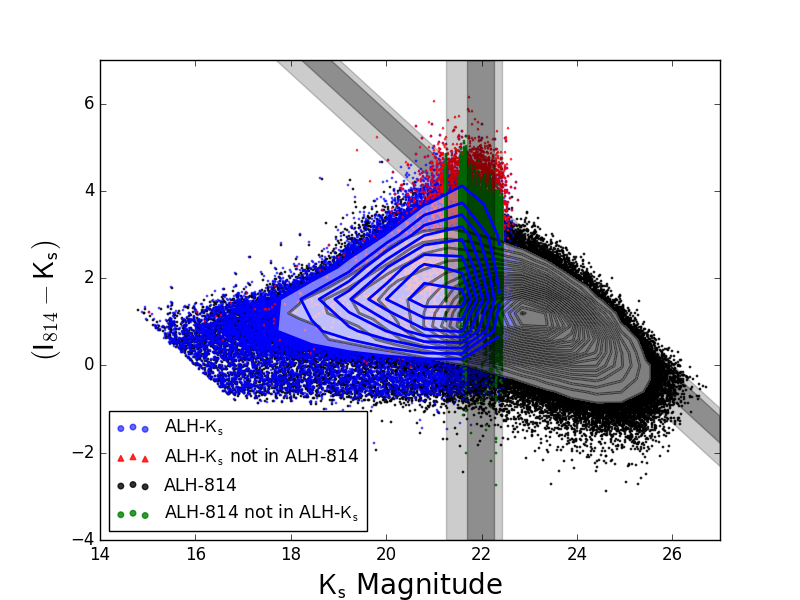}
\caption{Colour-magnitude diagrams. The left panel shows the one corresponding to the ALHAMBRA-4 field alone, and the right panel shows the full catalogue. Black points (and contours) are objects in the ALHAMBRA F814W-band selected catalogue, while blue points (and contours) come from this work. The red points correspond to sources detected in the $K_s$-band image that have no counterpart in the  optical selected catalogue. Conversely, the green points mark objects in the F814W catalogue with no $K_s$ band flux detected. Grey bands mark the $K_s$ (vertical) and the $I_{814}$ (diagonal) $3\sigma$ magnitude limits. As each CCD has different properties, we represent them using a shadowed band spanning the range from the minimum to the maximum value. In the right panel the band  is darker in the central 68\% of the CCD magnitude limit values.} \label{fig34RL}
\end{figure*}

Figure \ref{fig34RL} shows the $(I_{814}-K_s)$ vs $K_s$ colour-magnitude diagram for the ALHAMBRA-4 field in the left panel, and for the whole ALHAMBRA $K_s$-band catalogue in the right panel. We present both, so that the reader can see a cleaner case with fewer points and more homogeneous data and magnitude limits (the single ALHAMBRA-4 field), as well as the diagram for the whole catalogue.  The shadowed bands represent the magnitude limits for the  $K_s$ images (vertical) and the F814W images (diagonal), and their width is due to the inhomogeneity of the achieved magnitude limits. Black dots and contours correspond to the ALHAMBRA catalogue ($I_{814}$-band selected), while the blue dots and contours correspond to the  $K_s$-band selected catalogue. We signal with red points the sources detected in the latter with no counterpart in the M14 catalogue. We have detected 503 new sources in the ALHAMBRA-4 field alone, and  a total of 4305 new sources in the full $K_s$-band catalogue. This diagram can be directly compared with the one in Figure \ref{colcoltheo}, and shows that this new catalogue does indeed fill the area that corresponds to moderate-redshift, high-luminosity, intrinsically red sources.  

%%%

\subsection{Tests of photometric calibration}

There are two obvious tests that we can perform to check the quality of our photometry: a first, basic test will be to link the photometry that we are measuring with the one previously published in the ALHAMBRA catalogue. Our catalogue, being based on a shallower image, includes only $\sim 20\%$ of the targets that ALHAMBRA includes over the same area, and uses image-defined apertures that can be significantly different, particularly in the case of targets which are faint in one or both of the detection images. However, over the common sample and in particular for bright objects, the photometry must be fully consistent. A second test will imply comparison with the aforementioned UltraVISTA catalogue, that overlaps a large part of our ALHAMBRA-4/COSMOS field and reaches $\sim2$ magnitudes deeper.

\subsubsection{Comparison with M14}

We have cross-matched our ALHAMBRA $K_s$-selected catalogue with the main ALHAMBRA catalogue published in M14, which was selected using a synthetic F814W image for detection. The combined catalogue includes a total of 89,877 sources (77,568 of them galaxies) for which we have 23-band photometry measured with different apertures in each of our catalogues. 

\begin{figure*}
\centering
\includegraphics[width=0.45\textwidth]{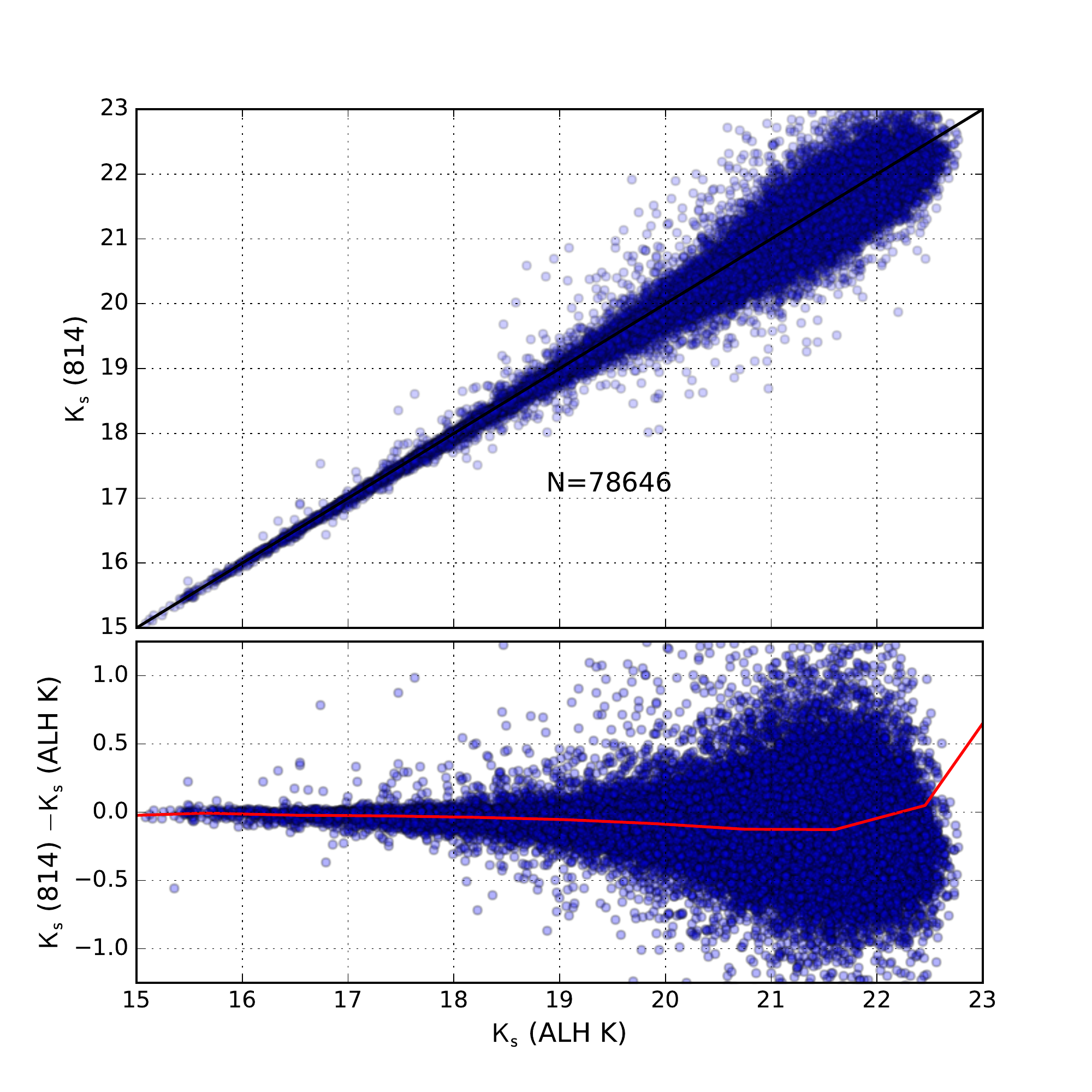}
\includegraphics[width=0.45\textwidth]{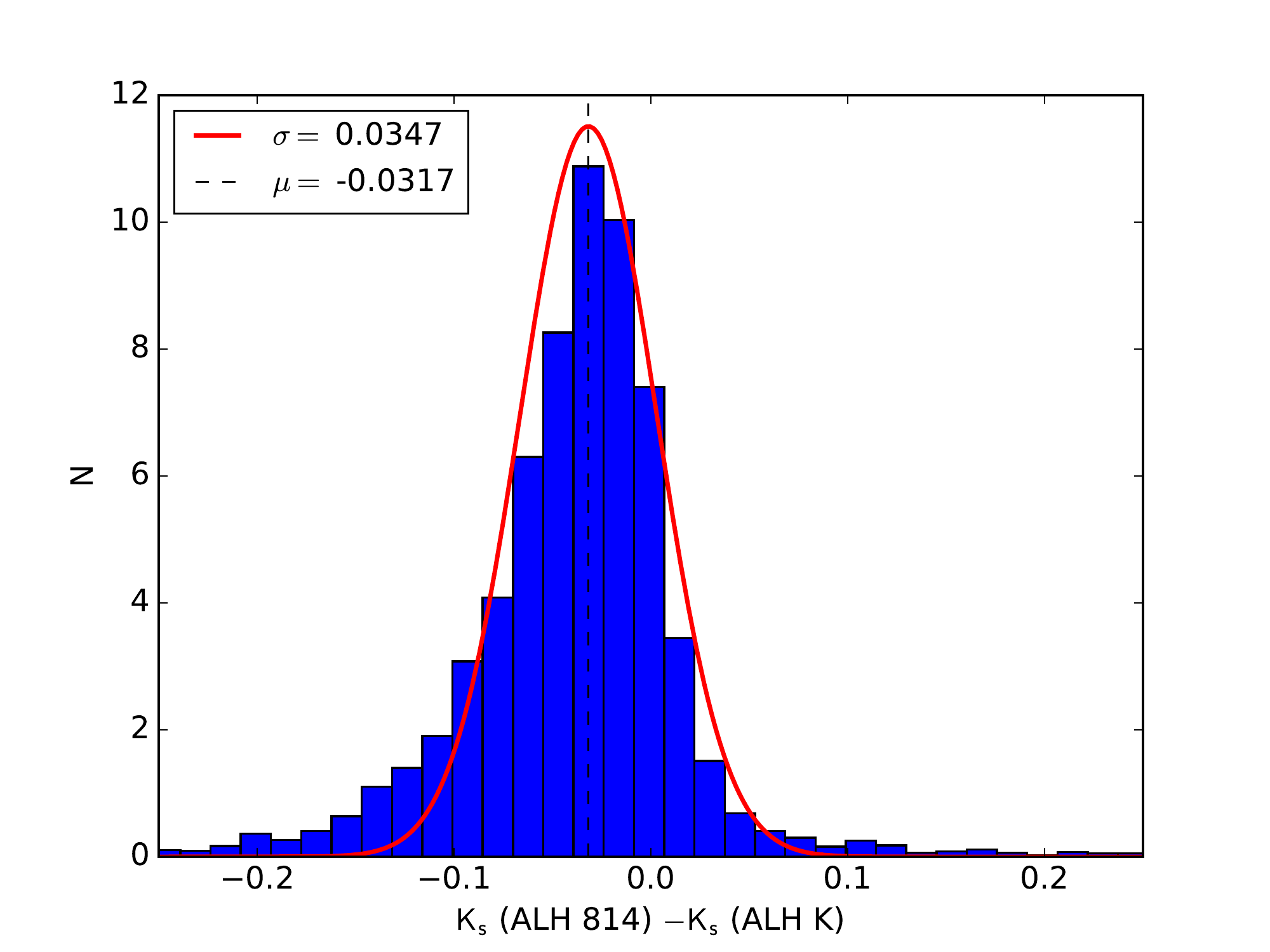}
\caption{Left panel: ALHAMBRA F814W-band catalogue and ALHAMBRA $K_s$-band catalogue photometry comparison. The right panel shows the distribution of $K_s$ magnitude differences for the bright sample (15<$K_s$<19), over-plotted with its Gaussian best fit whose parameters are given in the inset.}
\label{alhalhf814}
\end{figure*}

We have compared the $K_s$ band photometry of each object in this work with the one in M14, and show the result in Figure \ref{alhalhf814}. As expected, there is hardly any observable bias in the comparison for the bright sources ($K_s$<19), and the net average difference is comparable to or smaller than the typical photometric uncertainty. The right-hand panel in Figure \ref{alhalhf814} shows the distribution of the magnitude differences for bright sources, whose median is $\approx 0.03$ magnitudes. This value indicates that the $K_s$ magnitudes in the original ALHAMBRA catalogue are (in average) slightly brighter than the ones we obtain. We have tested that this effect is caused by the fact that the apertures defined by the F814W image are larger than the ones defined by the $K_s$ band, which pushes for a slightly larger flux to be measured in them\footnote{We must remark, however, that this effect is strongly intertwined with another effect which pushes in the opposite direction: in general, the apertures that we use are, by definition, better suited to measure the $K_s$-band flux, which is thus expected to be slightly larger in our measurement. This effect tends to be more noticeable for the faintest objects.}. We must insist that, in any case, both the scatter and the typical photometric uncertainties at the faint end of the catalogue are larger than this average effect.

\subsubsection{Comparison with UltraVISTA}

We perform a second, external consistency check of our photometry comparing it with the already mentioned UltraVISTA catalogue, which was observed in the same band \citep{2013ApJS..206....8M}. The data we compare correspond to the overlap between ALHAMBRA-4 and the UltraVISTA COSMOS field. As we discussed in Section \ref{stargal}, the total overlapping area is $\sim 0.21 \mathrm{deg^2}$ and the number of sources in common is 9,579. 

\begin{figure*}
\centering
\includegraphics[width=0.45\textwidth]{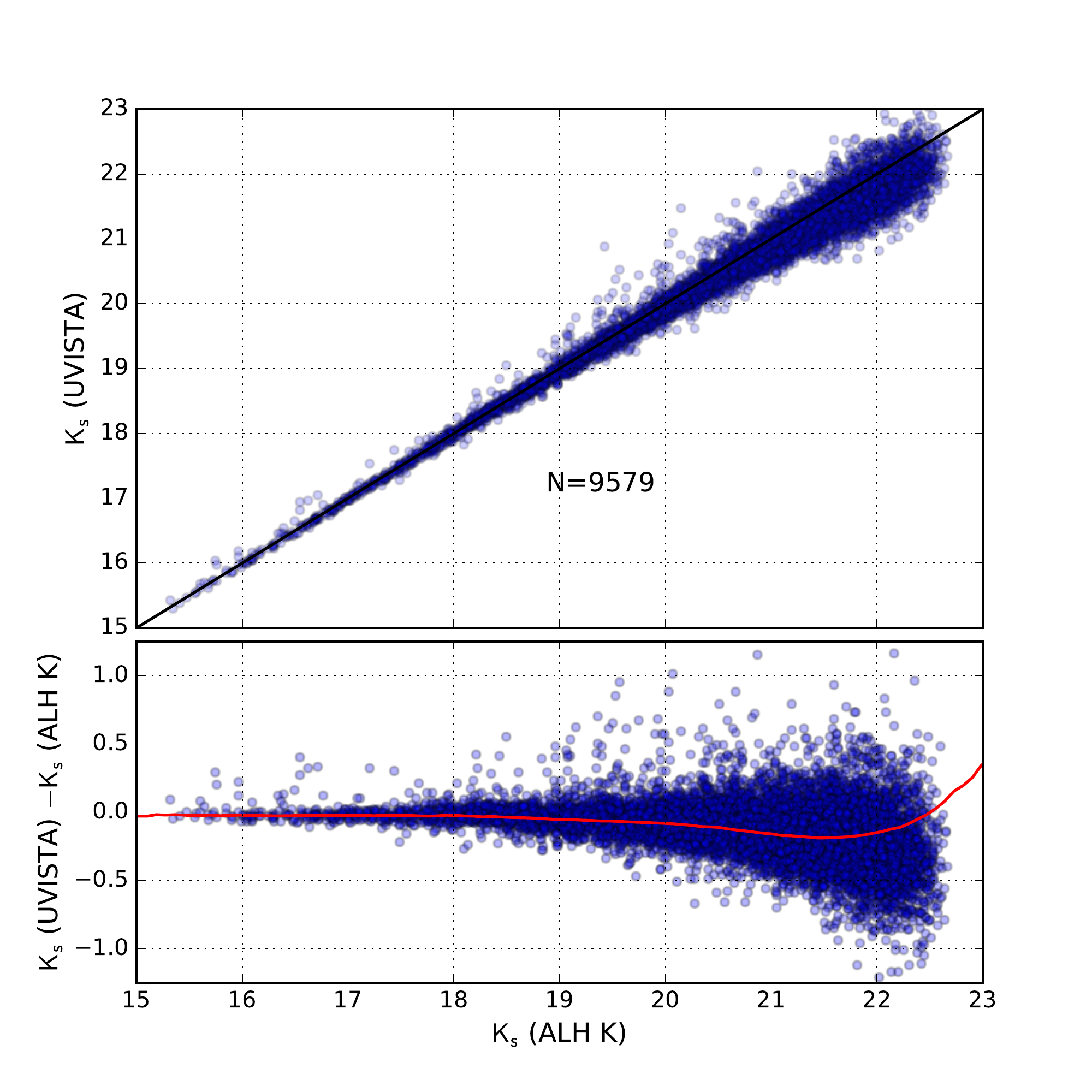}
\includegraphics[width=0.45\textwidth]{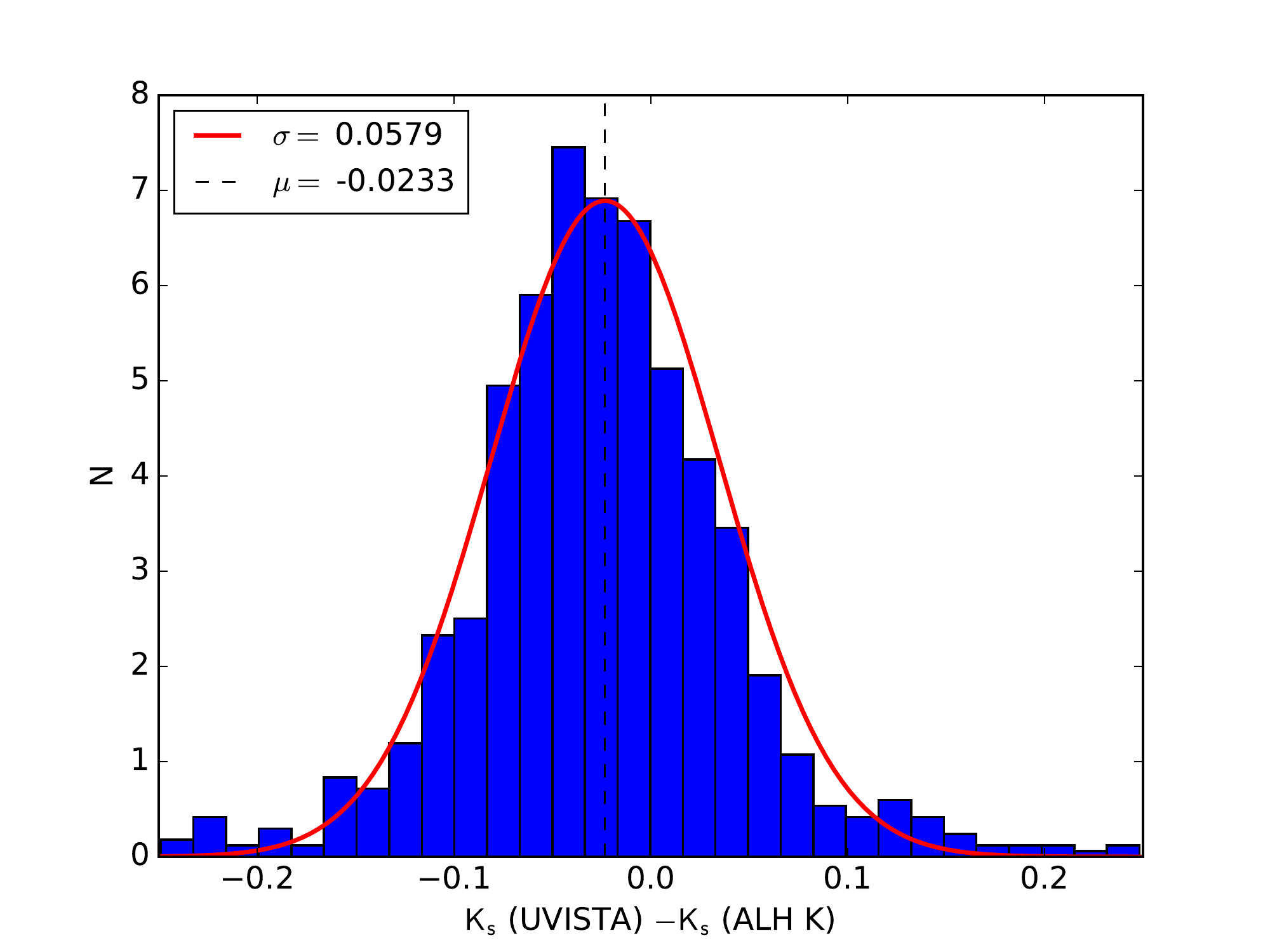}
\caption{Left panel: UltraVISTA and ALHAMBRA $K_s$ band catalogue photometry comparison for the ALHAMBRA-4 COSMOS field. The right panel shows the distribution of magnitude differences for the bright sample (15<$K_s$<19), over-plotted with its Gaussian best fit whose parameters are given in the inset.}\label{alhuvista}
\end{figure*}

In the left panel of Figure \ref{alhuvista} we show the results of the comparison of the $K_s$-band magnitudes for the objects in the common sample. As UltraVISTA is deeper than ALHAMBRA we can check our photometry all the way down to the ALHAMBRA $K_{s}$ magnitude limit. 

Selecting only bright targets $(15.5<K_s<19)$ to avoid the larger photometric uncertainties at the faint end, we can confirm an excellent agreement between both datasets. The right panel of Figure \ref{alhuvista} proves this result: we  find a systematic difference of 0.02 magnitudes---which is, in fact, comparable with the calibration uncertainty of the UltraVISTA data compared to the COSMOS catalogues and 2MASS \citep{2012A&A...544A.156M}. For the fainter sample the scatter between both datasets becomes larger, but remains always within the typical photometric uncertainties of both catalogues.

\subsection{Photometric Redshift Accuracy}

Once we have tested the correctness of the photometry performed on our images we can proceed to check the quality of the photometric redshifts, which are one of the key ingredients of our catalogue. As we did in the previous section for the photometry, we will perform two separate tests: an external one, comparing our photometric redshifts with those compiled from spectroscopic catalogues covering the same areas, and an internal one, comparing our results with the ones originally published in M14, whose quality was already assessed in that work.

In what follows we will use, in order to assess the quality of the photometric redshifts, the normalized median absolute deviation $\sigma_{\rm NMAD}$, as defined in \cite{2006A&A...457..841I}:
\begin{equation}
\mathrm{\sigma_{NMAD}=1.48 \times median} \left(\frac{\left| \delta z - \mathrm{median}(\delta z )\right|}{1+z_s} \right),
\label{sigmanmad}
\end{equation}
where $z_{\rm s}$ is the spectroscopic redshift and $\delta z = (z_{\rm s} - z_{\rm b})$ is the difference between the spectroscopic and the Bayesian photometric values. This parameter allows an accurate estimate of the rms for a Gaussian distribution and is less sensitive to outliers than the standard deviation. We will define the outlier rate (fraction of catastrophic errors) using two different criteria as in M14: $\eta_1$ is the fraction of sources that verify $\frac{| \delta z| }{1+z_{\rm s}} >0.2 $ and ${\eta_2}$ represents the fraction of sources that verify $\frac{| \delta z| }{1+z_{\rm s}} >5 \times \sigma_{\rm NMAD}$.

One of the features of BPZ \citep{2000ApJ...536..571B} is that it can be forced to use the information in a spectroscopic redshift sample to re-calibrate photometric zero points in each band. To do this the program compares the observed photometry with the one that would be expected of the galaxy templates at the known (spectroscopic) redshift for each object. If this comparison shows a significant zero-point bias in a given filter, this value is added to all the magnitudes in that filter and the whole process is iterated. \cite{2014MNRAS.441.2891M} discussed in detail this photometric redshift-based zero-point re-calibration. We have  used this feature for our $K_s$-band catalogue, finding very small corrections (median absolute deviation per filter $\approx 0.02$ mags), and a small but noticeable improvement in the quality of the photometric redshifts.

\subsubsection{Spectroscopic redshift comparison} 

We have repeatedly mentioned that one of the advantages of the ALHAMBRA Survey is the overlap with other well-known fields. This allowed M14 to compile a sample of 7144 galaxies with spectroscopic redshifts from public databases. We have identified the objects in this spectroscopic sample within our catalogue in order to compare the spectroscopic redshifts, $z_{\rm s}$ with the bayesian photometric redshifts estimated in this work, $z_{\rm b}$.

%%%%%%%%%%%%%%%%%%%%%%%%%%%%%%%%%%%%%%%%%%%
\begin{figure}   
\centering
\includegraphics[width=\columnwidth]{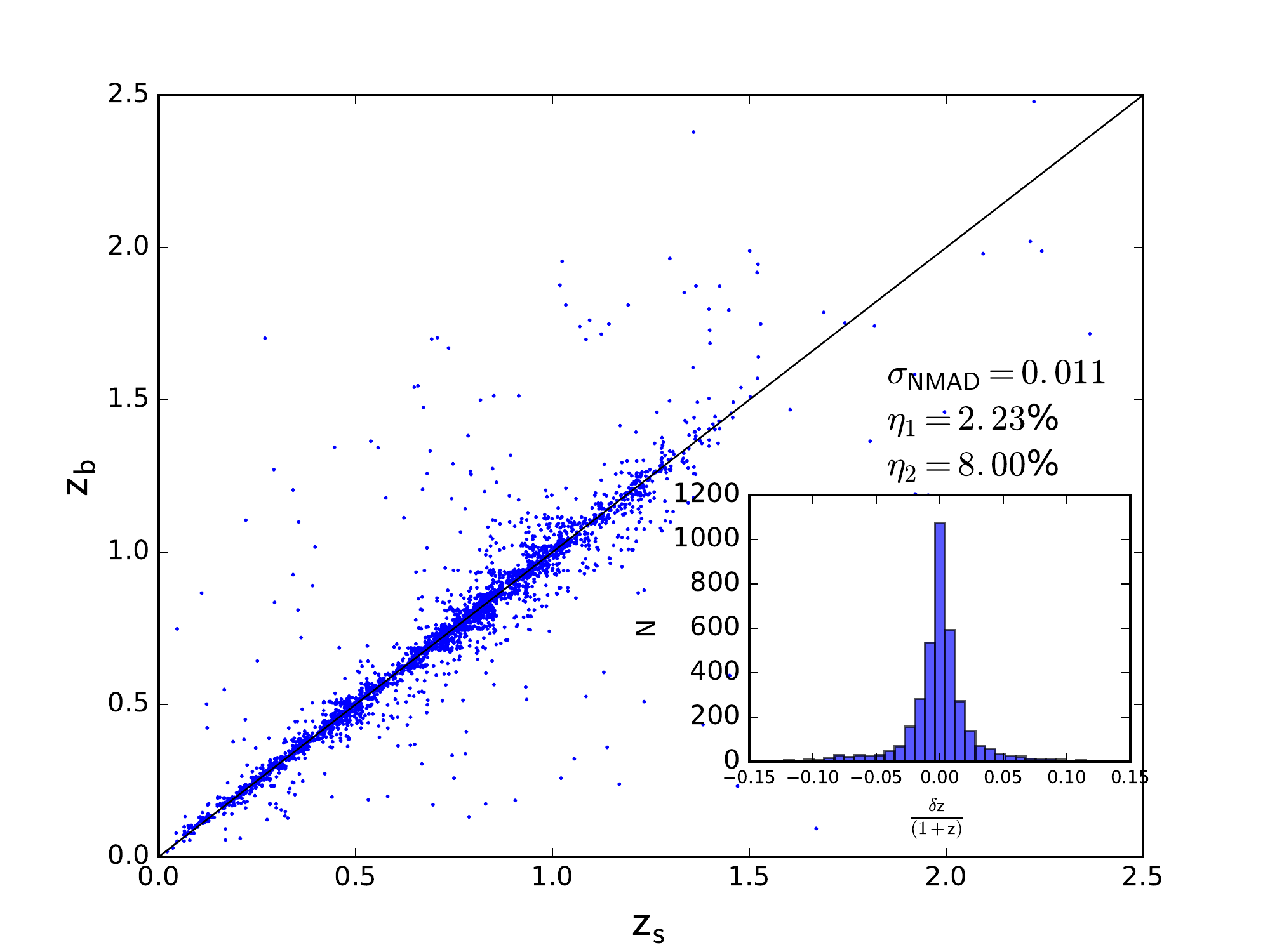}
\caption{ALHAMBRA $K_s$ photometric redshift $z_{\rm b}$ vs. spectroscopic redshift $z_{\rm s}$ for 3736 sources. The inset shows the distribution of the deviations $\delta_z/(1+z)$. The measured scatter is $\sigma_{\rm NMAD}=0.011$, with catastrophic error rates $\eta_1\sim 2.3\%, \eta_2\sim 8.0\%$.} \label{zpzs_1}
\end{figure}
%%%%%%%%%%%%%%%%%%%%%%%%%%%%%%%%%%%%%%%%%%%

We show in Figure \ref{zpzs_1} the result of the comparison of the photometric and spectroscopic redshifts, which in our case includes 3736 sources. The median magnitudes of the spectroscopic sample are $K_s=20.47$ and $I_{814}=21.96$, with respective first and third quartiles (19.62,21.24) and (21.11,22.66). We obtain a dispersion $\sigma_{\rm NMAD}=0.011$\footnote{The figure was $\sigma_{\rm NMAD}=0.015$ before the zero-point recalibration performed with the spectroscopic redshift sample, as explained in the previous paragraph.} for the total sample, and a catastrophic error rate $\eta_{1}\sim 2.3$\%. Both figures are similar to the ones obtained by M14 for their F814W $\leq 22.3$ sample.  

\begin{figure}
\centering
\includegraphics[width=\columnwidth]{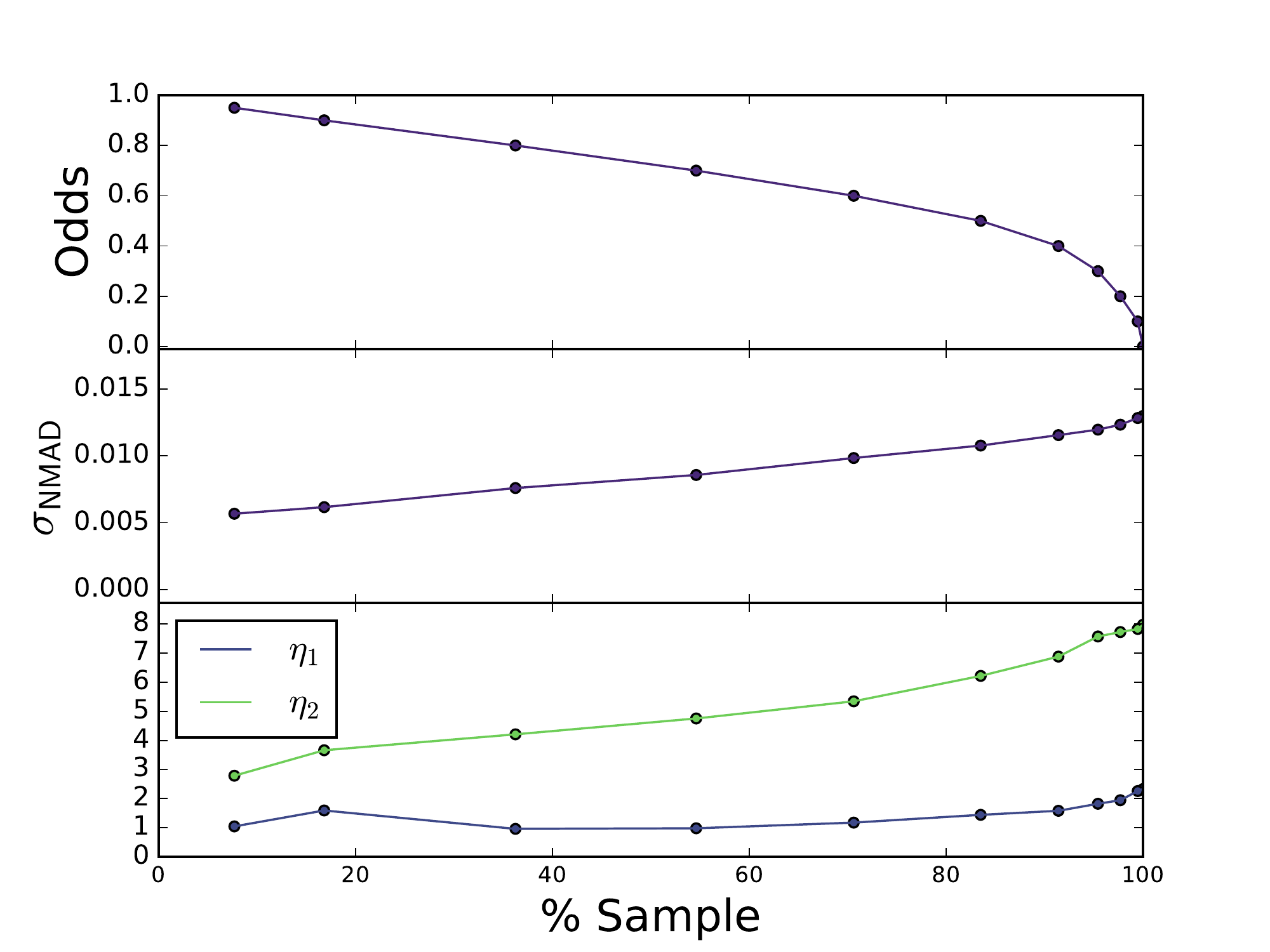}
\caption{Effect of the selection of sub-samples based on the {\it Odds} parameter. Top panel:Evolution of the {\it Odds} parameter with sample size. Middle panel: Evolution of the photometric redshift accuracy $\sigma_{\rm NMAD}$ with sample size, thresholded by {\it Odds} values. Bottom panel: Outlier rate parameters $\eta_1$ and $\eta_2$ as a function of sample size.} \label{oddseta}
\end{figure}

As we have mentioned in Section \ref{photoz}, among the output of the BPZ code we get for each object a Bayesian {\it Odds} parameter, which measures the affidability of the measured photometric redshift \citep{2000ApJ...536..571B}. We thus expect that both the dispersion $\sigma_{\rm NMAD}$ and the outlier rate parameters $\eta_1$ and $\eta_2$ should decrease when samples with increasingly large {\it Odds} values are selected. This effect is clearly shown in Figure \ref{oddseta}: all the quality indicators ($\sigma_{\rm NMAD}, \eta_1,\eta_2$) consistently improve when we impose a lower limit on the {\it Odds} parameter (top panel), at the obvious price of a decreasing sample size. In the extreme case, when only objects with {\it Odds}>0.95 are selected, the scatter falls to $\sigma_{\rm NMAD}=0.005$, but the sample size is less than 10\% of the original.

%%%%%%%%%%%%%%%%%%%%%%%%%%%%%%%%%%%%%%%%%%%
%%%%%%%%%%%%%%%%%%%%%%%%%%%%%%%%%%%%%%%%%%%

 \subsubsection{ALHAMBRA F814W catalogue photometric redshift comparison}
 
As shown in the previous section, the number of sources with spectroscopic information is scarce. In order to compile a larger sample with which our results can be compared, we have used the ALHAMBRA photometric redshifts calculated in M14 for the F814W-based catalogue. It is obvious that we are, after all, using the same imaging data for the same sources (those which are common to both catalogues), so we should necessarily reach similar results. However, we see this test as a necessary trial of the detection, aperture definition, and photometry processes we have performed.

In order to create a pseudo-spectroscopic sample, where the systematic effects will not be dominated by the photometric uncertainties, we have selected sources with magnitudes $K_s<19.5$ and $I_{814}<21$. From this sample we have excluded objects identified as stars in any of the two catalogues. This sample includes 10251 sources.

Comparing the photometric redshifts in M14 and in this work we obtain a scatter $\sigma_{NMAD}=0.009$, and catastrophic error rates  $\eta_1\sim 0.58\%$ and $\eta_2\sim 5.94\%$. As expected, the comparison between our catalogue and the ALHAMBRA F814W sample yields results that are much better than the spectroscopic comparison, even though the number of sources included in the analysis is larger. 

After this final check we are satisfied that our catalogue can be scientifically exploited. Figure \ref{SEDsamples} shows the photometry and best fit results for three example objects, which cover a wide range in photometric redshift, best-fitting spectral type and $K_s$ magnitude.

%%%%%%%%%%%%%%%%%%%%%%%%%%%%%%%%%%%%%%%%%%% 
\begin{figure}   
\centering
\includegraphics[width=\columnwidth]{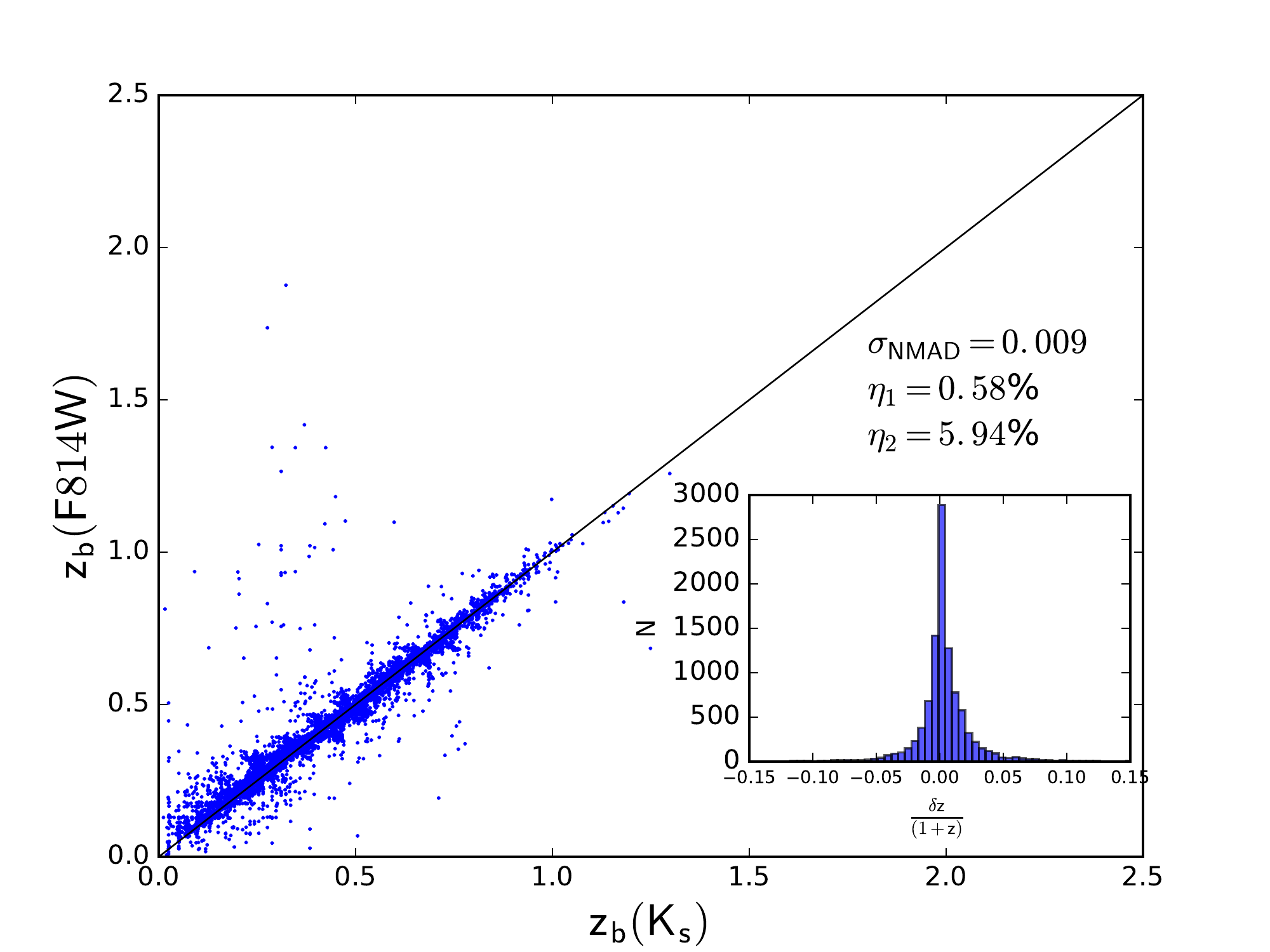}
\caption{$K_s$-band photometric redshift $z_{\rm b}(K_s)$ vs. photometric redshift $z_{\rm b}$(F814W) of 10,251 sources. In the inset we present the distribution of $\delta_z/(1+z)$. We measure $\sigma_{\rm NMAD}=0.009$ and a catastrophic error $\eta_1 \sim 0.58$\%. The small concentration of outliers at $z_{\rm b}(K_s)$ is due to the colour degeneracy between low-redshift red galaxies and moderate-redshift bluer ones.} \label{zpzs_2}
\end{figure}
%%%%%%%%%%%%%%%%%%%%%%%%%%%%%%%%%%%%%%%%%%%
%%%
 
%%%%%%%%%%%%%%%%%%%%%%%%%%
\begin{figure}
\centering
\includegraphics[width=0.45\textwidth]{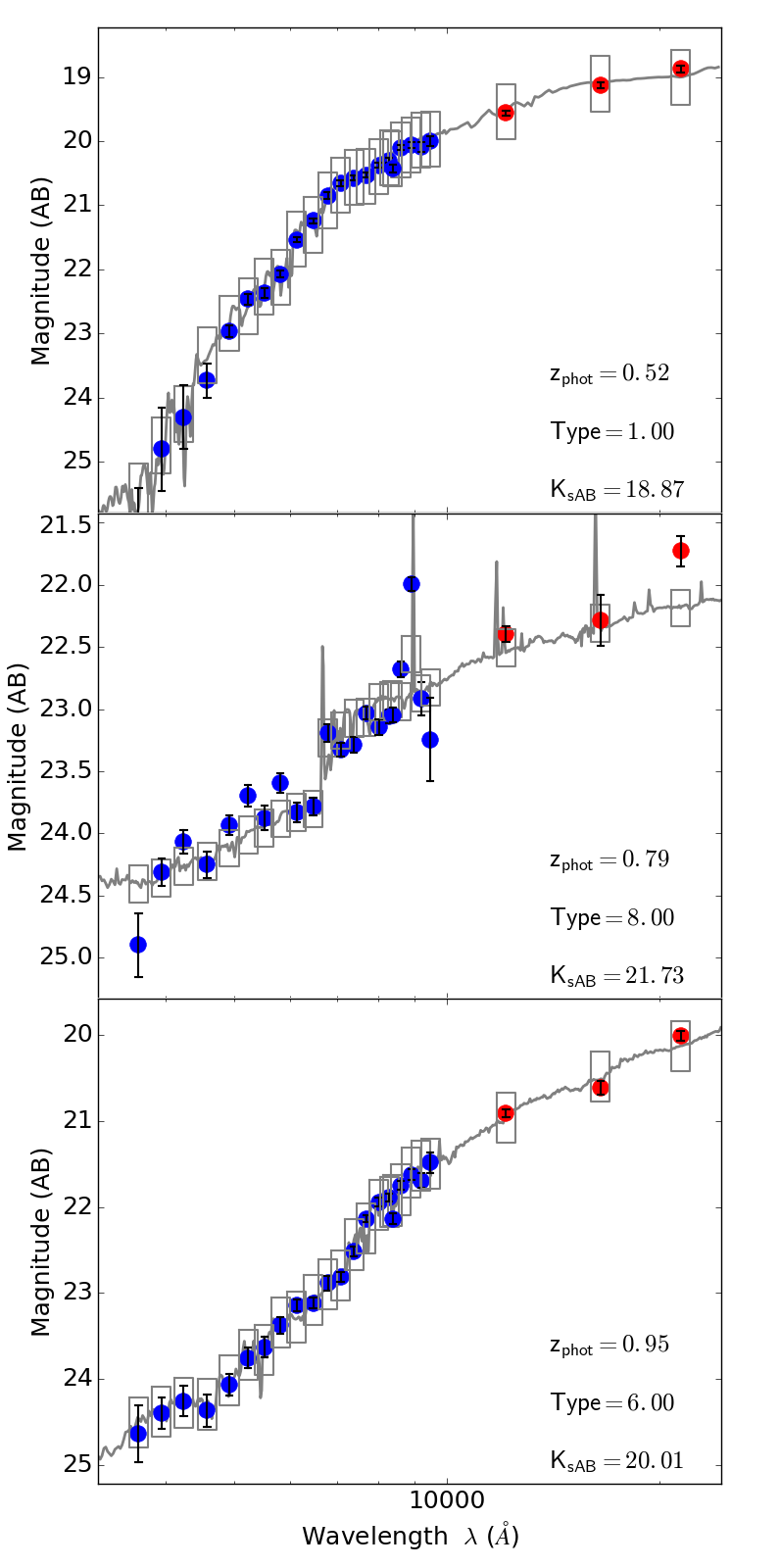}
\caption{Three examples of spectral fits to the ALHAMBRA data of different galaxies. Redshifts range from 0.5 to 1, SEDs from $T_B=1$ (elliptical) to $T_B=8$ (starburst), and $K_s$-band magnitudes from $\sim 19$ to $\sim 22$.  On each plot the grey line is the spectrum  corresponding to the best fit, as indicated in the inset text, and the grey rectangles are the model photometry in the 23+1 ALHAMBRA bands (we include the synthetic F814W image flux). We use blue (red) markers for the optical (near infrared) spectral range.} \label{SEDsamples}
\end{figure} 
%%%%%%%%%%%%%%%%%%%%%%%%%%

\subsection{Photometric Redshift Distribution}

After checking the quality of our catalogue regarding photometry and redshifts, and once we are satisfied that the number count distribution is correct, we can move on to the exploitation of the catalogue. The most basic functions to analyse include the individual and multivariate redshift-magnitude-spectral type distributions.

\begin{figure}
\centering
\includegraphics[width=\columnwidth]{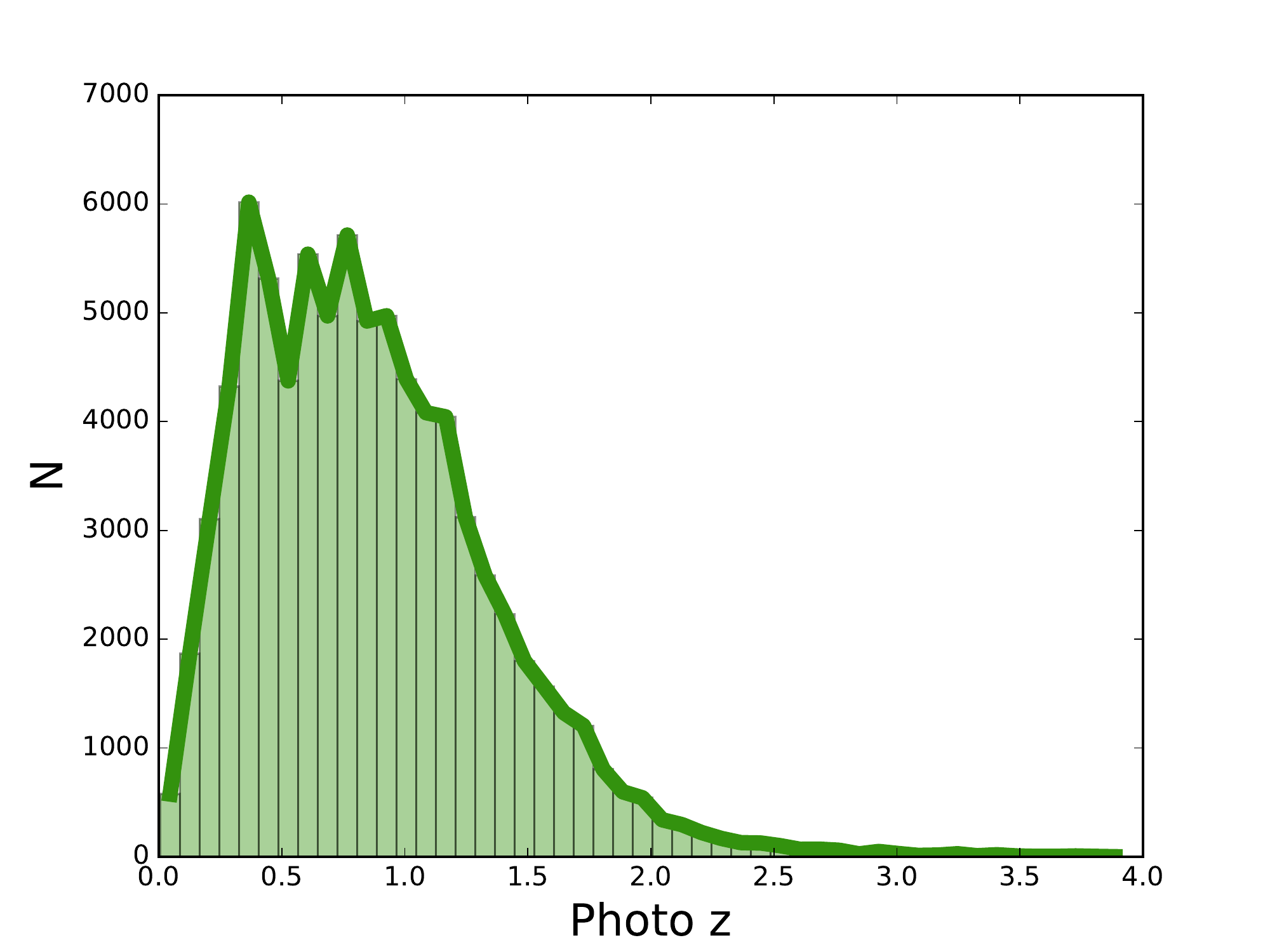}
\caption{Histogram of BPZ best-fitting photometric redshifts for a total of 81,931 galaxies.} \label{plotsPhotozNumber}
\end{figure}

In what follows we have cleaned our galaxy sample using the \texttt{COLOR\_CLASS\_STAR} parameter to avoid stars, leaving us with a catalogue of 81,873 galaxies. We show the histogram of best-fitting photometric redshifts for this sample in Figure \ref{plotsPhotozNumber}. The median redshift is $\left\langle z \right\rangle =0.80$, and the values of the first and third quartiles of the redshift distribution are $z_{\rm Q1}=0.47$, $z_{\rm Q3}=1.15$. 

%%%%%%%%%%%%%%%%%%%%%%%%%%%%%%%%%%%%%%%%%%%
%%%%%%%%%%%%%%%%%%%%%%%%%%%%%%%%%%%%%%%%%%%

\begin{figure*}
\centering
\includegraphics[width=\columnwidth]{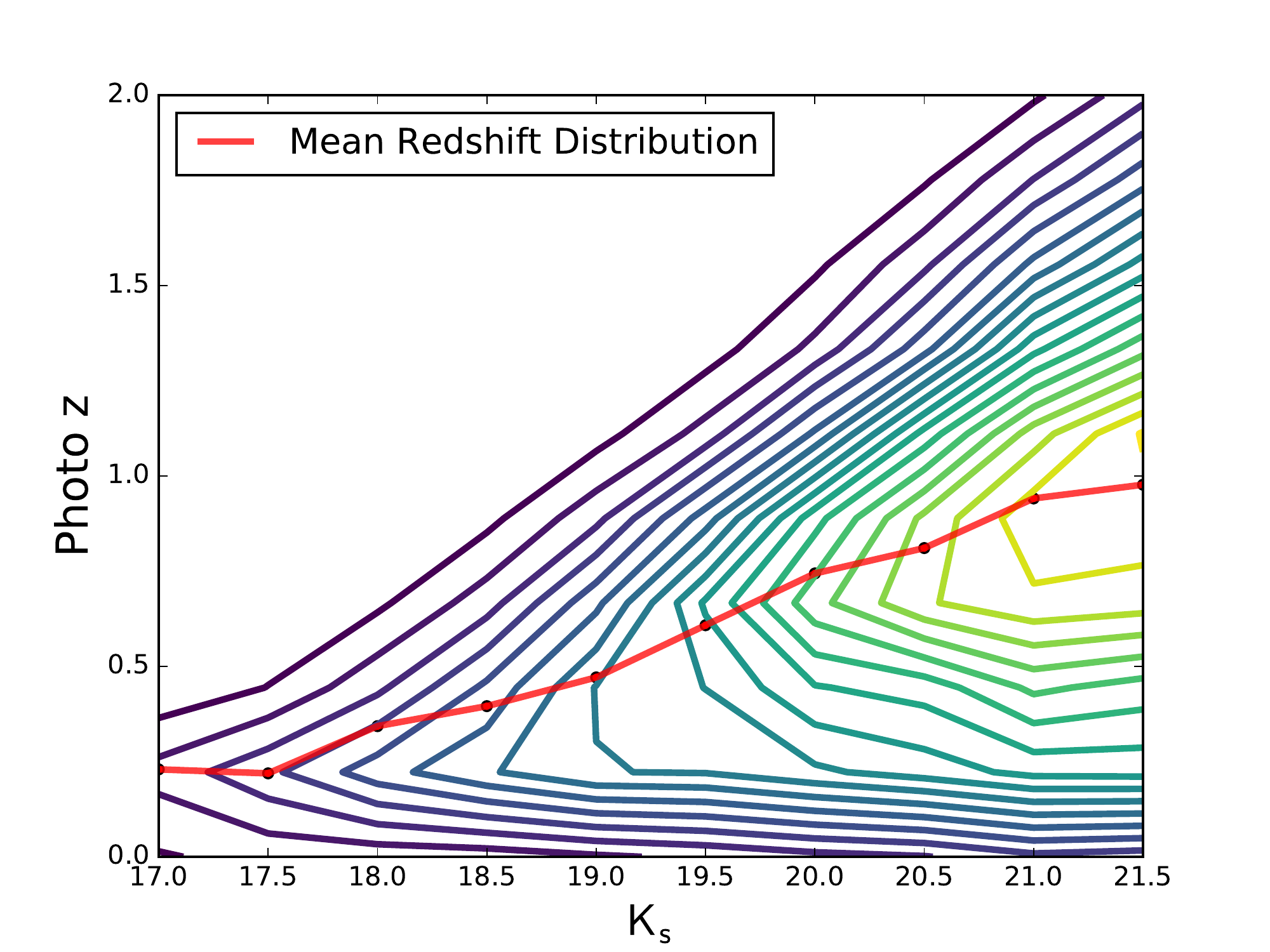}
\includegraphics[width=\columnwidth]{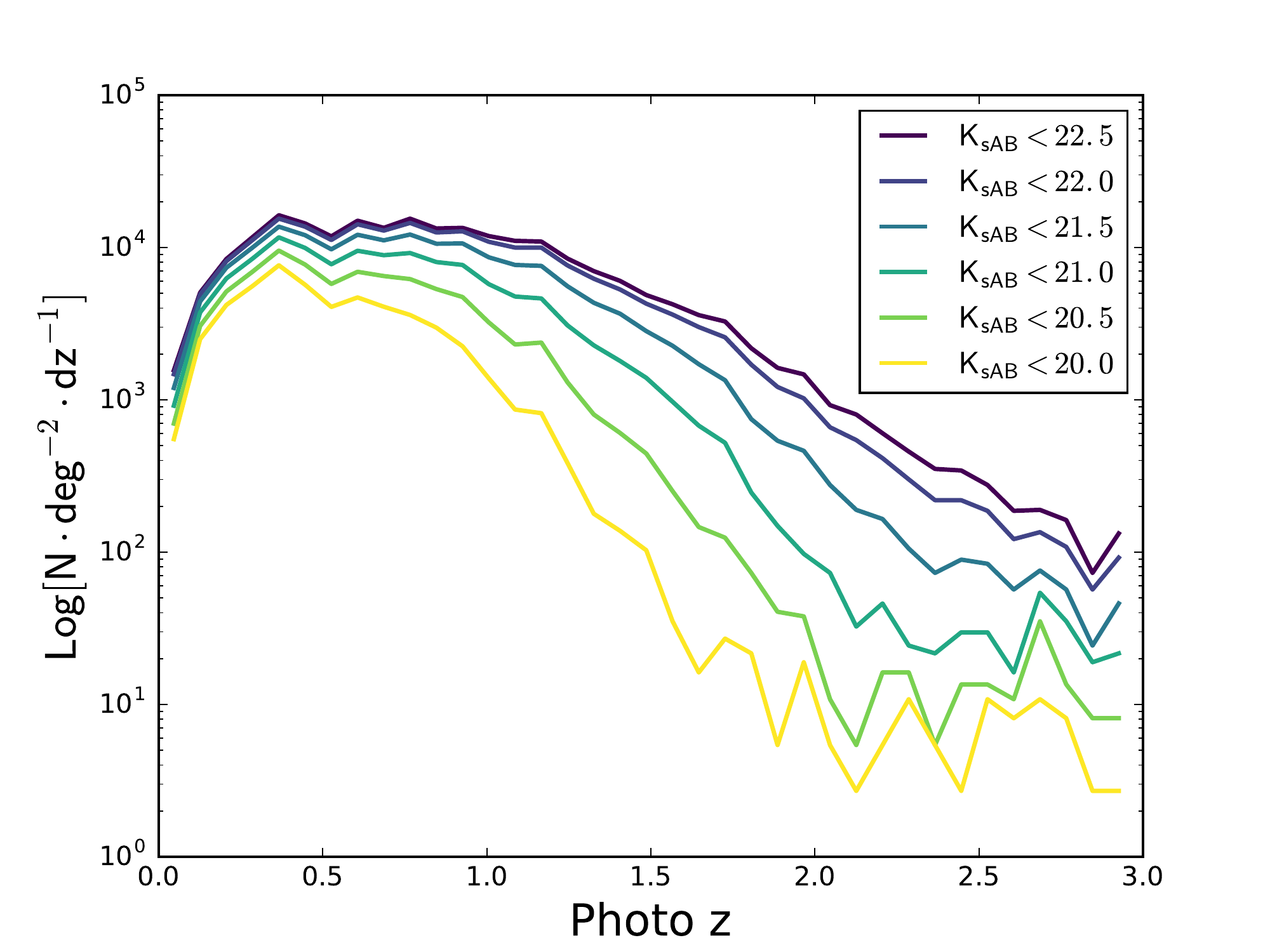} 
\caption{Left Panel: Density contour plot in the redshift-$K_s$ magnitude plane. The solid red line shows the evolution of the mean photometric redshift as a function of magnitude. Right Panel: Redshift distribution of galaxies in our catalogue, selected in successive $K_s$ magnitude cuts.}	\label{fig34RL_1}
\end{figure*}

The left panel of Figure \ref{fig34RL_1} shows a contour plot of the redshift-apparent magnitude plane. We have overplotted on it as a red line the evolution of the median redshift as a function of the $K_s$ magnitude. In the right panel of the same figure we present an alternative view of the same distribution, in this case showing the redshift distribution of galaxies for different apparent magnitude cuts. This plot shows very clearly that any increase in $K_s$-band depth represents a corresponding increase in redshift depth, as we expected from the colour-magnitude diagrams analysed in previous sections. The tail of the distribution towards high redshift is populated with intrinsically red objects, most of them showing very red $(I_{814W}-K_s)$ colours and in most cases absent from the general-purpose ALHAMBRA catalogue presented in M14. We will check this statement in the next section.

%%%%%%%%%%%%%%%%%%%%%%%%%%%%%%%%%%%%%%%%%%%
%%%%%%%%%%%%%%%%%%%%%%%%%%%%%%%%%%%%%%%%%%%

\subsection{Redshift distribution of galaxy types}

We will now focus on the analysis of the distribution of galaxy types in the catalogue. As was mentioned in the Introduction, the ALHAMBRA F814W catalogue shows a dearth of early type galaxies at redshifts $z \gtrsim 1.1$, which is induced by the passage of the 4000 \AA\ break and associated absorption at such redshift through the F814W filter, the one used for source detection by M14.

We will compare the redshift/SED distribution obtained by M14 (F814W-selected) with the one we have obtained with our catalogue selected in the $K_s$-band. This comparison will allow us to check whether we are, in fact, recovering those early-type galaxies at z$\gtrsim 1.1$.

\begin{figure}
\centering
\includegraphics[width=\columnwidth]{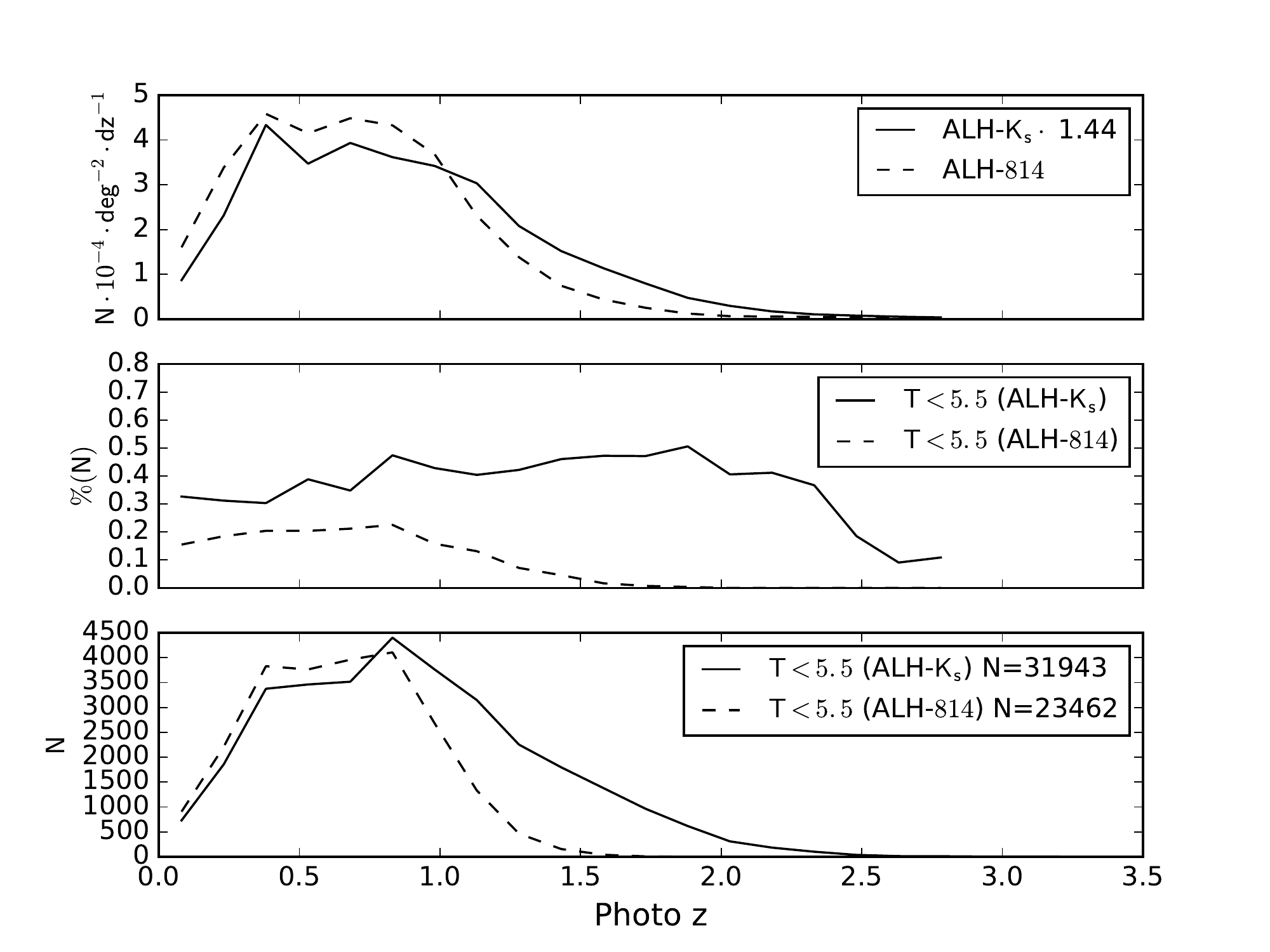}
\caption{Top: Photometric redshift distribution for the  $K_s$-selected (solid line) and F814W-selected (dashed line) samples. Centre: Fraction of galaxies in both samples whose best-fitting templates are 1<$T_B$<5.5, which we identify as early-type SEDs. Bottom: Absolute number of early-type sources at each redshift in each catalogue.} \label{galdensperc}
\end{figure}

Figure \ref{galdensperc} shows in the top panel the photometric redshift distribution of galaxies in the $K_s$ (solid line) and F814W (dashed line) catalogues. As the NIR sample is shallower we have applied a cut in the original ALHAMBRA catalogue ($I_{814}<23.5$), and to render both curves directly comparable we have multiplied the NIR distribution by a factor 1.44, so that the areas under both curves are the same. There is a hint of structure in both lines over the range $0.3<z<0.8$, which can be due to the large-scale structure which is obviously common to both catalogues.

It is remarkable that, even though the $K_s$-band selected sample presented here is less deep than the original ALHAMBRA catalogue, its tail extending to high redshift is clearly more noticeable. This is exactly what we expect from the recovery of $z\gtrsim 1$ early-type galaxies. To confirm this point we have measured the fraction of early-type galaxies (defined as those with $1 \leq T_B \leq 5.5$) at each redshift. This fraction is plotted in the lower panel of Figure \ref{galdensperc}. The increase in the fraction of early-type galaxies is significant already at low redshift ($\approx 35\%$ at $z<1$ in the NIR sample, compared to $\approx 20\%$ in the F814W case). But the effect is much stronger at higher redshifts---the fraction of early-type galaxies decreases to reach almost zero at redshift $z\approx 1.5$ in the ALHAMBRA M14 catalogue, whereas we still observe a sizeable fraction of early types ($\approx 40-50\%$) out to the highest redshifts accesible to our catalogue ($z\approx2.5$).  The middle panel on Figure \ref{galdensperc} shows the fraction of early-type galaxies in both ALHAMBRA catalogues. The apparent plateau in the fraction of early-type galaxies in our catalogue between redshifts $z \sim 0$ and $z \sim 2.5$ is due to the combination of an approximately flux-selected sample and the  increasing efficiency of the red-galaxy selection with redshift.

%%%%%%%%%%%%%%%%%%%%%%%%%%%%%%%%%%%%%%%%%%% 
%%%%%%%%%%%%%%%%%%%%%%%%%%%%%%%%%%%%%%%%%%%

\section{Discussion}  

Once we have presented the procedure leading to the construction of the $K_s$-band selected ALHAMBRA catalogue and checked the consistency and quality of the photometry and photometric redshift catalogue, we finish by introducing the catalogue itself and presenting a brief discussion of some of the possibilities that the catalogue offers for the future.

\subsection{Catalogue description}

The ALHAMBRA $K_s$-selected catalogue is distributed as a single compressed file, including individual files from all 48 individual pointings\footnote{In the survey webpages {\tt http://www.alhambrasurvey.com} and {\tt http://cloud.iaa.csic.es/alhambra/}}.  It is also accessible through the Virtual Observatory through its Spanish site \footnote{{\tt http://svo.cab.inta-csic.es}}. Each individual catalogue, corresponding to a single pointing (one CCD area) includes a header that documents the column information. 

We include in Appendix \ref{appcolumns} a list with the content and description of the individual columns in each file. 

\subsection{IRAC cross-match data}

As we have shown in the previous sections, the ALHAMBRA survey  overlaps with fields that have been extensively studied by other projects. This will allow us to benefit from ancillary information added to our catalogue, as was the case with the spectroscopic sample. 

One of the most interesting additions for the particular aims addressed in this work is the possibility of extending the photometry further into the infrared range, using public catalogues provided by different teams. Although for some of our targets and objectives we will also benefit from data from the all-sky WISE Survey \citep{2010AJ....140.1868W} and the {\it Spitzer} MIPS instrument, we will concentrate here only in the data from the {\it Spitzer Space Telescope} Infra-Red Array Camera (IRAC, \citealt{2004ApJS..154...10F}). Images taken with this instrument will add photometric data in four new bands, centered at 3.6, 4.5, 5.8, and $8.0 \mu$. 

There are several public fields where deep IRAC data have been taken and analysed. The extra information provided by these images will be crucial for some of our targets: we must keep in mind that the main objective of our work is to detect and analyse objects with very red intrinsic $(I_{814}-K_s)$ colours, which in some cases implies that we will only have solid detections of their flux in the $JHK_s$ filters, combined with strong limits on their flux in the visible range. The extension to the $3-8 \mu$ wavelength range provided by IRAC means that we will be able to observe a much wider rest-frame spectral window, including the characteristic potential downturn in the flux of early-type galaxies at wavelengths $\lambda > 2.5 \mu$ in the rest frame.

\begin{figure*}
\centering
\includegraphics[width=0.32\textwidth]{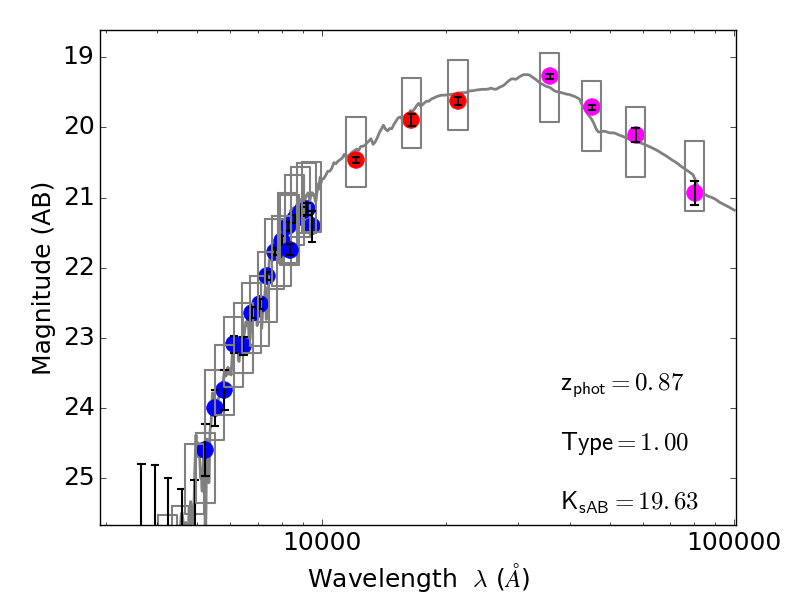}
\includegraphics[width=0.32\textwidth]{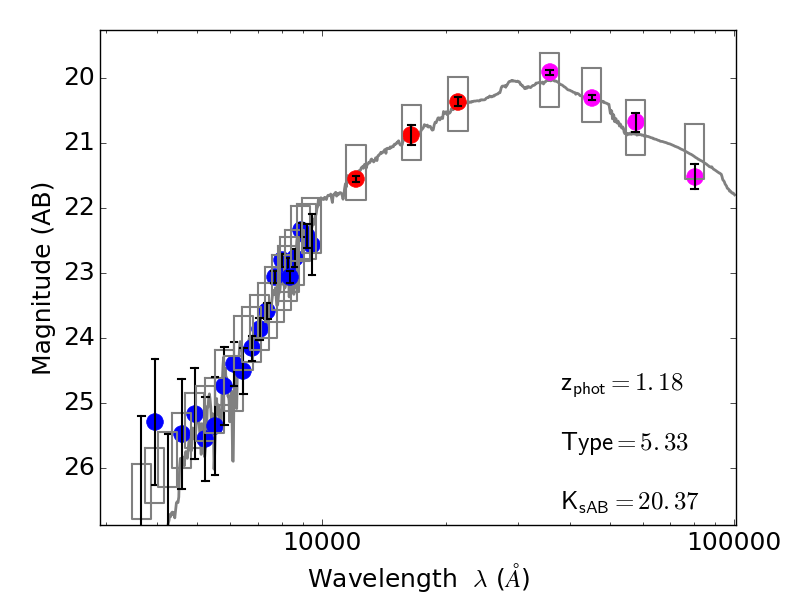}
\includegraphics[width=0.32\textwidth]{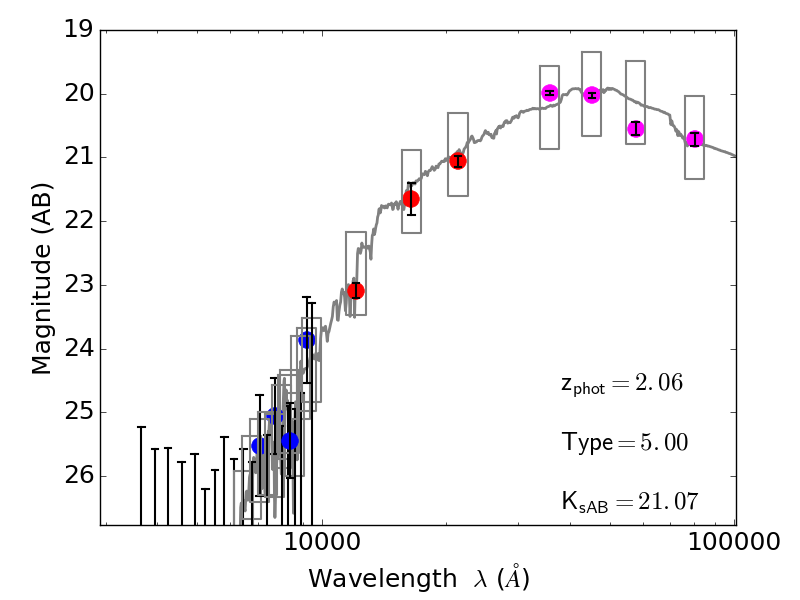}
\includegraphics[width=0.32\textwidth]{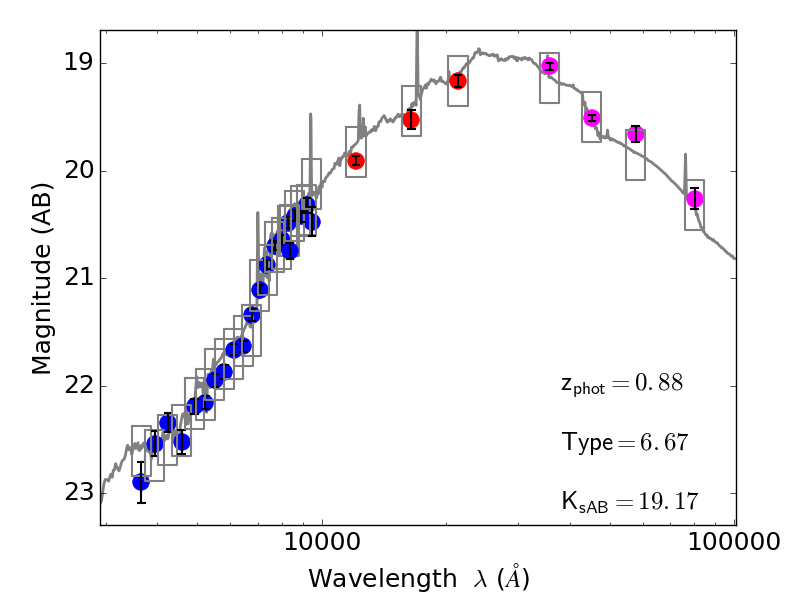}
\includegraphics[width=0.32\textwidth]{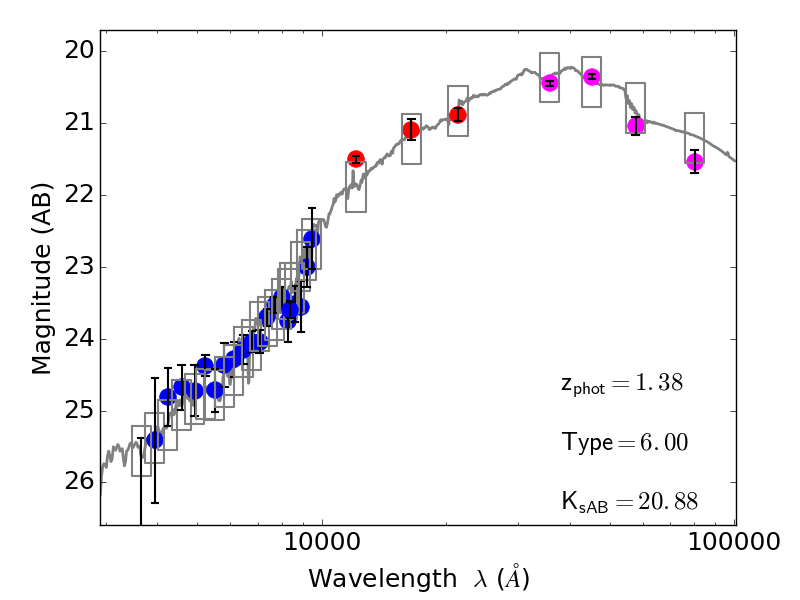}
\includegraphics[width=0.32\textwidth]{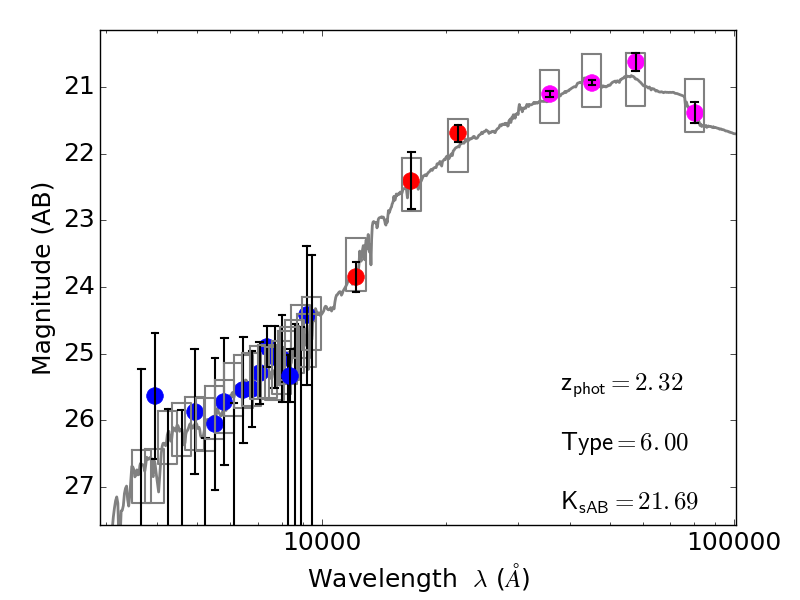}
\includegraphics[width=0.32\textwidth]{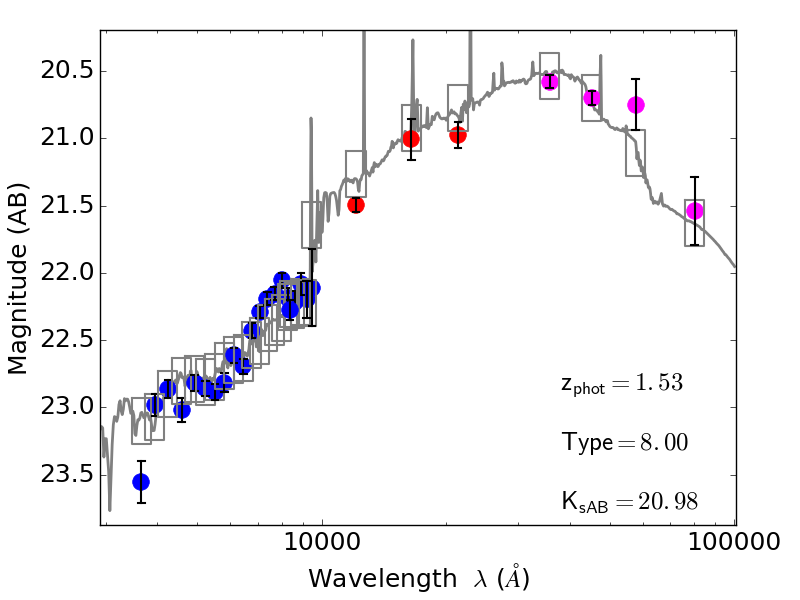}
\includegraphics[width=0.32\textwidth]{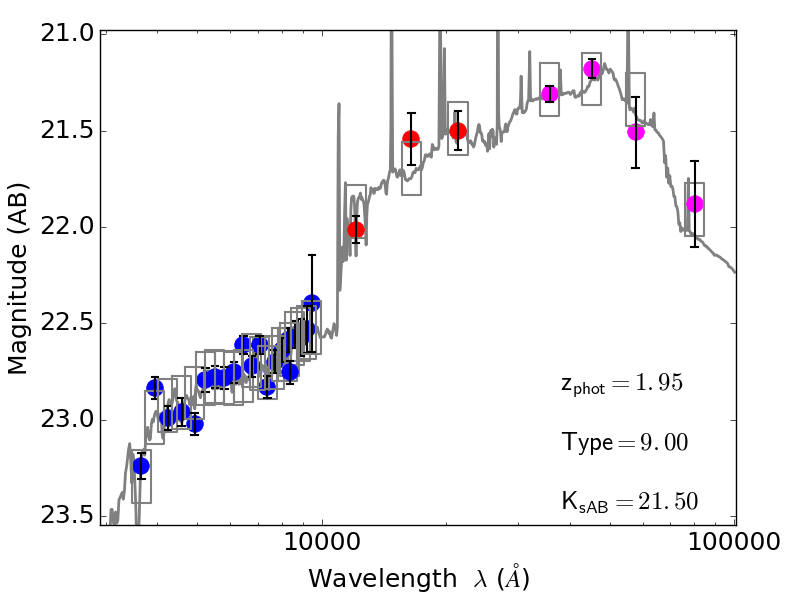}
\includegraphics[width=0.32\textwidth]{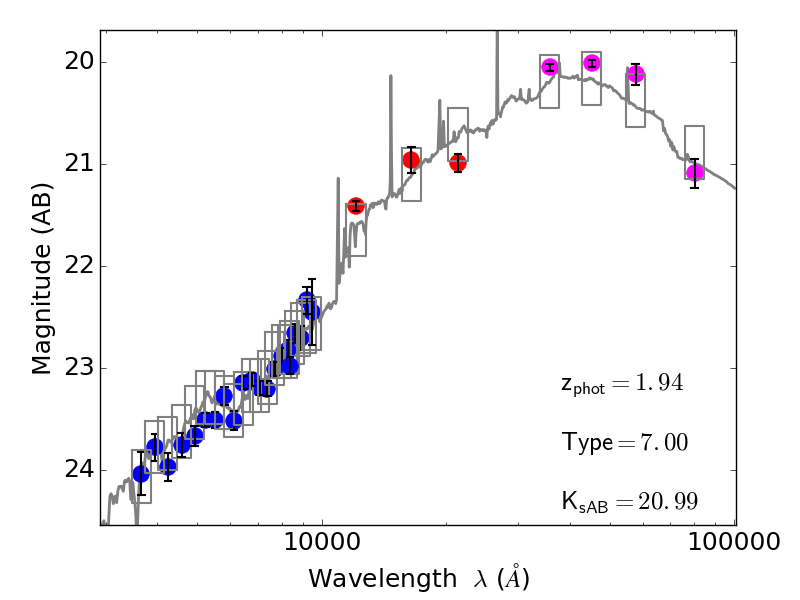}
\caption{Example ALHAMBRA+IRAC pseudospectra and best fits for a sample of galaxies. The top (medium, bottom) row shows spectral data corresponding to early-type (late-type, star-forming) SEDs, and in each case the redshift grows from left to right. In all panels the blue markers correspond to ALHAMBRA data in the visible range, red to ALHAMBRA $JHK_s$, and magenta to {\it Spitzer} IRAC data.} \label{SED1s}
\end{figure*}

As a test in this first stage we have cross-matched our catalogue in the ALHAMBRA-7 field (ELAIS-N1, \citealt{1999ESASP.427.1011R}) with the public data of the {\it Spitzer} Wide-area Infrared Extragalactic Survey \citep[SWIRE, ][]{2003PASP..115..897L}. Over 75\% (11,756/15,262) of our ALHAMBRA-7 $K_s$-selected sources have counterparts in the IRAC database. We have plotted the SEDs of some of these galaxies in Figure \ref{SED1s}, covering a wide range of galaxy types (top to bottom) and redshifts (left to right). A glance at this figure is enough to show the complementarity between the ALHAMBRA visible and NIR data and the extension allowed by the IRAC available observations.

\begin{figure*}
\centering
\includegraphics[width=0.48\textwidth]{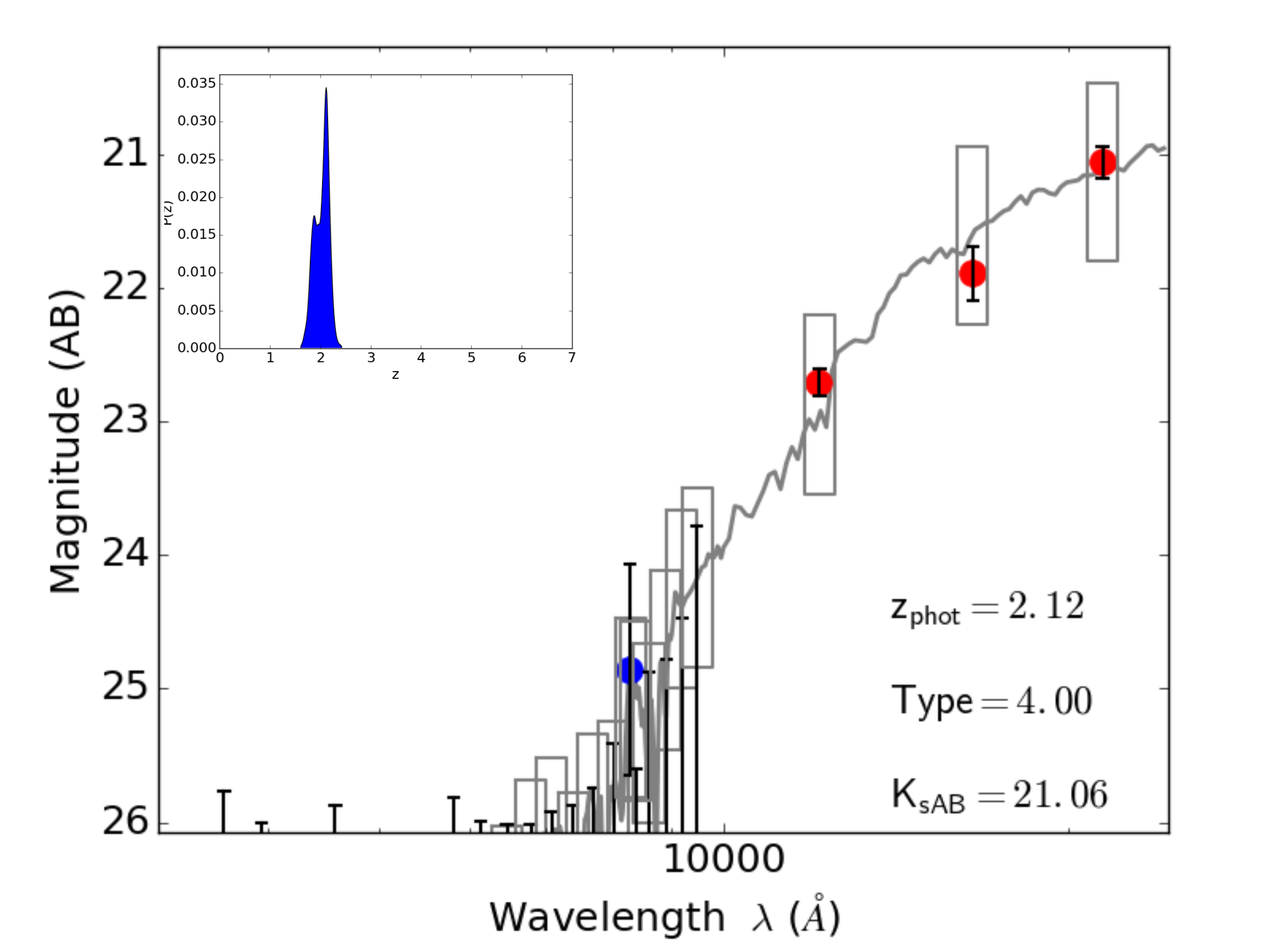}
\includegraphics[width=0.48\textwidth]{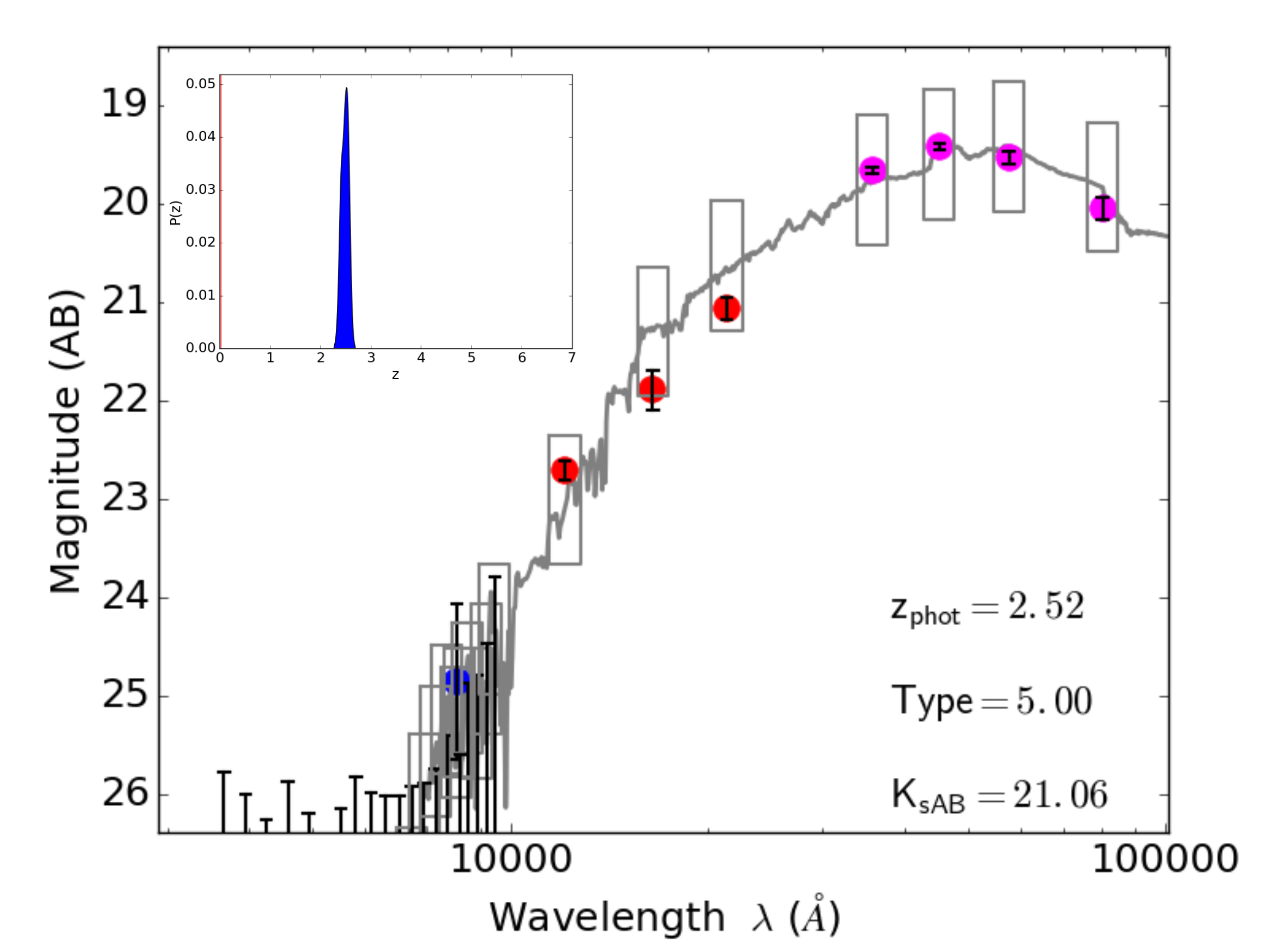}
\caption{Pseudospectrum and best-fitting SED of an Extremely Red Object from our catalogue. The left panel shows the fit obtained using the ALHAMBRA 20+3 filter data set. The right panel shows the result of the analysis of the same object, once the IRAC data are added. In each case the inset panel shows the Bayesian redshift probability function $p(z)$.}\label{SEDs}
\end{figure*}

Within this ALHAMBRA-7 sample we have found a significant number of galaxies with extremely red colours, that render them observable in ALHAMBRA only in the $JHK_s$ filters. For these objects the IRAC data becomes crucial to allow for a robust characterization. We have detected in IRAC 225 of 246 such sources, all with colour $(I_{814}-K_s)$>4. In Figure \ref{SEDs} we show in the left panel our best fit for one such galaxy, whereas the right panel shows how the fit changes and the results improve when the IRAC data are added.  A careful look at the $p(z)$ insets shows that the functions almost do not overlap, which may seem incorrect. However, as explained in \cite{2002MNRAS.330..889F}, in these cases where only a very limited amount of data points are available for the fit, the systematic errors induced by the lack of coverage in the SED template library can be dominant, and would render the $p(z)$ curves wider and compatible. In the next Section we will give some figures about the sample of this kind of objects that we can extract from our catalogue. Galaxies of this kind are interesting by themselves, and deserve a more detailed analysis which will be the target of a forthcoming manuscript (Nieves-Seoane {\it et al.}, in preparation). 

Other fields where similarly deep IRAC observations are available include ALHAMBRA-2 (DEEP2), ALHAMBRA-4 (COSMOS), ALHAMBRA-5 (HDFN), and ALHAMBRA-6 (GROTH). A full combination and analysis of this dataset will be the objective of a future manuscript (Nieves-Seoane {\it et al.}, in preparation).

%%%%%%%%%%%%%%%%%%%%%%%%%%%%%%%%%%%%%%%%%%%
%%%%%%%%%%%%%%%%%%%%%%%%%%%%%%%%%%%%%%%%%%%

%%%%%%%%%%%%%%%%%%%%%%%%%%%%%%%%%%%%%%%%%%%
%%%%%%%%%%%%%%%%%%%%%%%%%%%%%%%%%%%%%%%%%%%

\subsection{Extremely Red Objects}

As discussed in the previous section, the ALHAMBRA  $K_s$-band catalogue includes a significant number of sources typically classified as Extremely Red Objects (EROs, \citealt{1988ApJ...331L..77E}). These objects are usually selected according to their very red colours (for example $(I-K)>4)$, and classified using both their visible and NIR colours as, for example, in the $BzK$ selection technique \citep{2004ApJ...617..746D}. Most of the EROs can be classified either as passively evolving or as dusty star-forming galaxies \citep{2002A&A...381L..68C}. 

\begin{figure}
\centering
\includegraphics[width=0.45\textwidth]{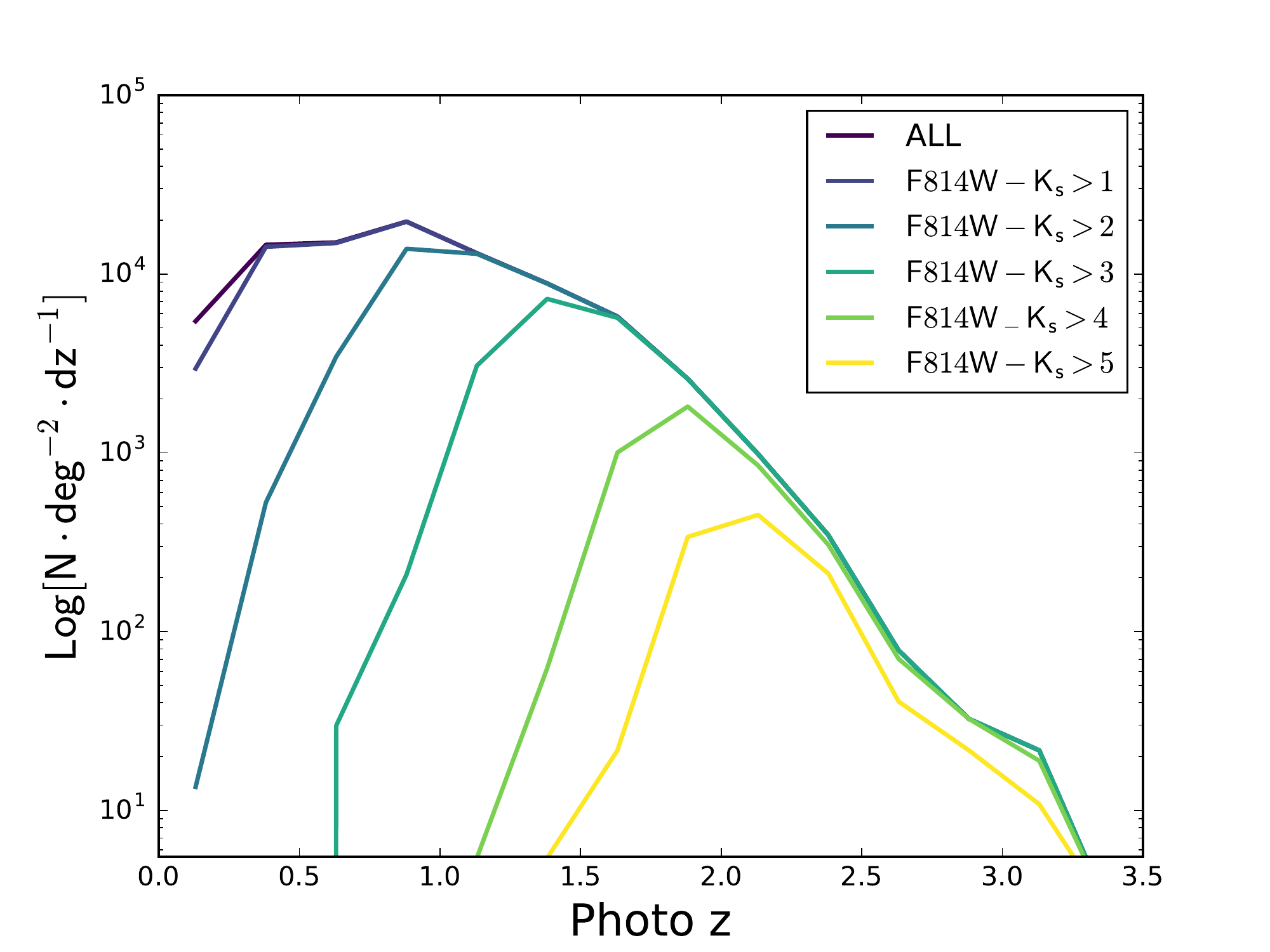}
\caption{Distribution of photometric redshifts for different $(I_{814}-K_s)$ colour-selected samples, for galaxies with $T_B<5.5$.}\label{galdistribcol}
\end{figure}

We can use our very deep synthetic F814W image, combined with the $K_s$ band images used for galaxy detection in our catalogue, to select EROs based on different $(I_{814}-K_s)$ thresholds. As was indicated above, these cuts in $(I_{814}-K_s)$, induce an almost one-to-one selection in redshift for galaxies characterized by early-type SEDs. This is clearly seen in Figure \ref{galdistribcol}, where the redshift distributions of galaxies with $T_B<5.5$ are plotted for different threshold values of $(I_{814}-K_s)$. We list in Table \ref{tabredshift} the sizes and values of the first quartile, median, and third quartile redshift of those samples. Taking a reference value for the threshold selection of $(I_{814}-K_s)>4$ the total number of such ERO candidates in our catalogue is 1539.

\begin{table}
\centering
\caption{Characteristics of the redshift distribution of samples of objects characterized by early-type SEDs ($T_B<5.5$) and different $(I_{814}-K_s)$ colour thresholds.}
\begin{tabular}{ccccc}
\hline
$(I_{814}-K_s)$ & N & $z_{\rm 1Q}$ & $z_{\rm med}$ & $z_{\rm 3Q}$\\ 
\hline 
All & 31943 & 0.54 & 0.85 & 1.19 \\
>1  & 30815 & 0.59 & 0.87 & 1.20 \\
>2  & 18278 & 0.93 & 1.14 & 1.43 \\
>3  &  7503 & 1.32 & 1.49 & 1.71 \\
>4  &  1539 & 1.75 & 1.89 & 2.06 \\
>5  &   408 & 1.95 & 2.11 & 2.27 \\ 
\hline
\end{tabular}
\label{tabredshift}
\end{table}

The colours of most of the galaxies in these ERO samples fit those of old, massive, passively evolving galaxies \footnote{We must remark, however, that BPZ does not fit the dust content as a separate parameter, but includes the effect of fixed amounts of dust within the templates themselves. Because of this, objects with a very high dust content may not be correctly identified in our catalogue.}. These galaxies are one of the key steps in galaxy evolution, and the study of their properties is essential for the understanding of the early phases of the evolution of elliptical galaxies. 

\begin{figure}
\centering
\includegraphics[width=0.45\textwidth]{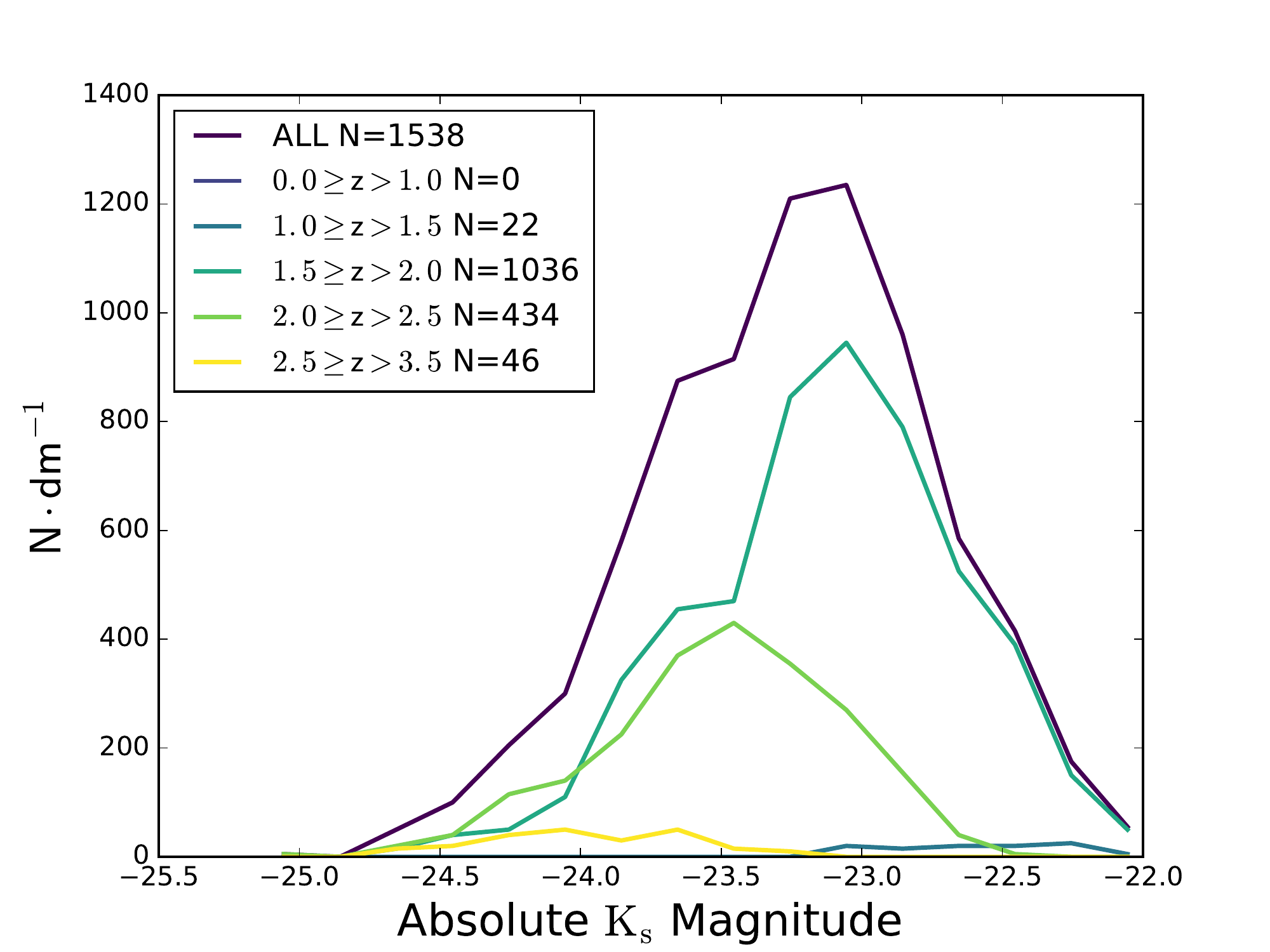}
\caption{Distribution of the $K_s$ absolute magnitudes in our catalogue for galaxies with $(I_{814}-K_s)>4$ and $T_B<5.5$, in different redshift ranges.} \label{magandmassdisV1}
\end{figure}

We show the distribution of $K_s$ absolute magnitudes for the sample of early-type galaxies in Figure \ref{magandmassdisV1}, where it is clear that the vast majority of them lie within the redshift interval $1.5<z<2.5$, reaching luminosities as high as $M_K \approx -24.5$. A more detailed study of the EROs classification from our catalogue will be presented in a future paper.

%%%%%%%%%%%%%%%%%%%%%%%%%%%%%%%%%%%%%%%%%%%
%%%%%%%%%%%%%%%%%%%%%%%%%%%%%%%%%%%%%%%%%%%

%%%%%%%%%%%%%%%%%%%%%%%%%%%%%%%%%%%%%%%%%%%
%%%%%%%%%%%%%%%%%%%%%%%%%%%%%%%%%%%%%%%%%%%

\section{Conclusions}

We have presented in this paper the photometric and photometric redshift catalogue of sources detected in the  ALHAMBRA $K_s$-band images. The catalogue includes photometry for 94,182 sources distributed over seven fields, covering a total area of 2.47 deg$^2$. This catalogue is different from the original ALHAMBRA catalogue presented in \citet{2014MNRAS.441.2891M} because that sample was selected based on a synthetic F814W image, similar to an $I$-band selection. Such a selection is biased against intrinsically red galaxies at redshift $z \gtrsim 1$, an effect that became noticeable in several of the recent works based on the ALHAMBRA survey. This issue sparked our interest in producing a new catalogue where this bias would be avoided by selecting in the reddest band available.

Source detection and photometry was performed using SExtractor in dual mode. We estimated the photometric errors using the method presented by \citet{2003AJ....125.1107L}, and used an adapted version of the masks created in \cite{2014MNRAS.441.1783A} to define the survey window. Star-galaxy separation was performed using a colour-colour diagram, and tested with the SEDs of the NGSL stellar library. We calculated detailed completeness functions for every pointing using the deeper UltraVISTA catalogue of the ALHAMBRA-4/COSMOS field as reference. We applied this completeness functions to extend our magnitude limit and the number counts to magnitude $K_s \approx 21.9$. Two separate tests were performed to check the photometric accuracy of our catalogue: an internal test against the photometry of the objects common to our catalogue and the original ALHAMBRA catalogue presented in M14, and an external test using the objects common to our catalogue and UltraVISTA. In both cases the cross-catalogue accuracy has been shown to be compatible with that expected from the respective uncertainties, and with no significant bias. 

We completed our catalogue by running the BPZ2.0 code over our sample, including the zeropoint photometric recalibration option that uses a spectroscopic redshift sample to refine, at the same time, both the photometry and the photometric redshift accuracy. Using a spectroscopic redshift sample with 3736 galaxies, and the normalized median absolute deviation (NMAD) as a estimator of the accuracy of our results, we obtain $\sigma_{\rm NMAD}=0.011$, and a catastrophic error rate $\eta_1 \sim 1.3\%$, both comparable to the ones obtained by \cite{2014MNRAS.441.2891M}. We performed a second comparison, in this case with the photometric redshifts in the ALHAMBRA F814W-selected catalogue. This comparison yields $\sigma_{\rm NMAD}=0.009$ with a catastrophic error rate  $\eta_1 \sim 0.58\%$.

As expected, because of the motivation of our work, the photometric redshift distribution segregated by galaxy type shows that many of the new $K_s$-selected sources fill the dearth of early-type galaxies in the F814W-selected sample at $z \gtrsim 1$. We will present, in forthcoming works, a detailed analysis of this population, including the combination of our data with catalogues covering other wavelengths, in particular the {\it Spitzer}-IRAC public catalogues that overlap several of the ALHAMBRA fields. We have presented in this work some examples of how these extra photometric bands add to the information content of our catalogue.

\section*{Acknowledgements}

This work is based on observations collected at the German Spanish Astronomical Center, Calar Alto, jointly operated by the Max-Planck-Institut f\"ur Astronomie (MPIA) and the Instituto de Astrof\'{\i}sica de Andaluc\'{\i}a (CSIC). This study makes use of data from AEGIS, a multiwavelength sky survey conducted with the Chandra, GALEX, Hubble, Keck, CFHT, MMT, Subaru, Palomar, Spitzer, VLA, and other telescopes and supported in part by the NSF, NASA, and the STFC. This research has made extensive use of NASA's Astrophysics Data System.

This work was mainly supported by the Spanish Ministry for Economy and Competitiveness and FEDER funds through grants AYA2010-22111-C03-02 and AYA2013-48623-C2-2, and by the Generalitat Valenciana through project PrometeoII 2014/060. We also acknowledge support from the Spanish Ministry for Economy and Competitiveness and FEDER funds through grants AYA2012-39620, AYA2013-40611-P, AYA2013-42227-P, AYA2013-43188-P, AYA2013-48623-C2-1, ESP2013-48274, AYA2014-58861-C3-1,  Junta de Andaluc\'{\i}a grants TIC114, JA2828, P10-FQM-6444, and Generalitat de Catalunya project SGR-1398. BA has received funding from the European Union's Horizon 2020 research and innovation programme under the Marie Sklodowska-Curie grant agreement No 656354. PTI acknowlegdes support from CONICYT-Chile grant FONDECYT 3140542.

%%%%%%%%%%%%%%%%%%%%%%%%%%%%%%%%%%%%%%%%%%%%%%%%%%



\bibliographystyle{mnras}
\bibliography{bibliografia.bib} % if your bibtex file is called example.bib

%%%%%%%%%%%%%%%%%%%%%%%%%%%%%%%%%%%%%%%%%%%%%%%%%%

%%%%%%%%%%%%%%%%% APPENDICES %%%%%%%%%%%%%%%%%%%%%

\appendix
\section{Estimation of the survey completeness}
\label{appcompleteness}

We will estimate the completeness fraction of our survey as a function of the $K_s$-band magnitude using the UltraVISTA data as reference. Only one of our fields (ALHAMBRA-4) overlaps with this survey, so we will calculate an accurate completeness function using the four pointings of this field and scaling the results to the rest of the survey.

The basic idea is to compare the number of sources detected in the common area by UltraVISTA and ALHAMBRA in each magnitude interval. We have fitted the usual Fermi function
\begin{equation}
F(m) = [1 + \exp ((m-m_{\rm{c}})/\Delta m)]^{-1}
\end{equation}
to the data in each of the four CCDs where we have observations from both surveys. The parameters $m_{\rm{c}}$ and $\Delta m$ correspond respectively to the 50\% completeness magnitude and to a measurement of the width of the decreasing part of the Fermi function. We remark that the fits were adjusted to the data themselves, with no binning of the data involved. 

In a first run we fitted each of the four CCDs separately, and obtained the values of $m_{\rm{c}}$ and $\Delta m$ in all four cases. We checked that the widths were compatible with each other--in all cases the value was close to $\Delta m = 0.3$. We also checked that the value of the completeness limit indicator $m_{\rm{c}}$ was strongly correlated to the nominal $5\sigma$ limit of each field, which allows us to use the latter as a proxy for the former. In particular, this will be crucial to extend our analysis to the fields which cannot be directly compared to UltraVISTA or other deeper surveys.

\begin{figure}
\centering
\includegraphics[width=\columnwidth]{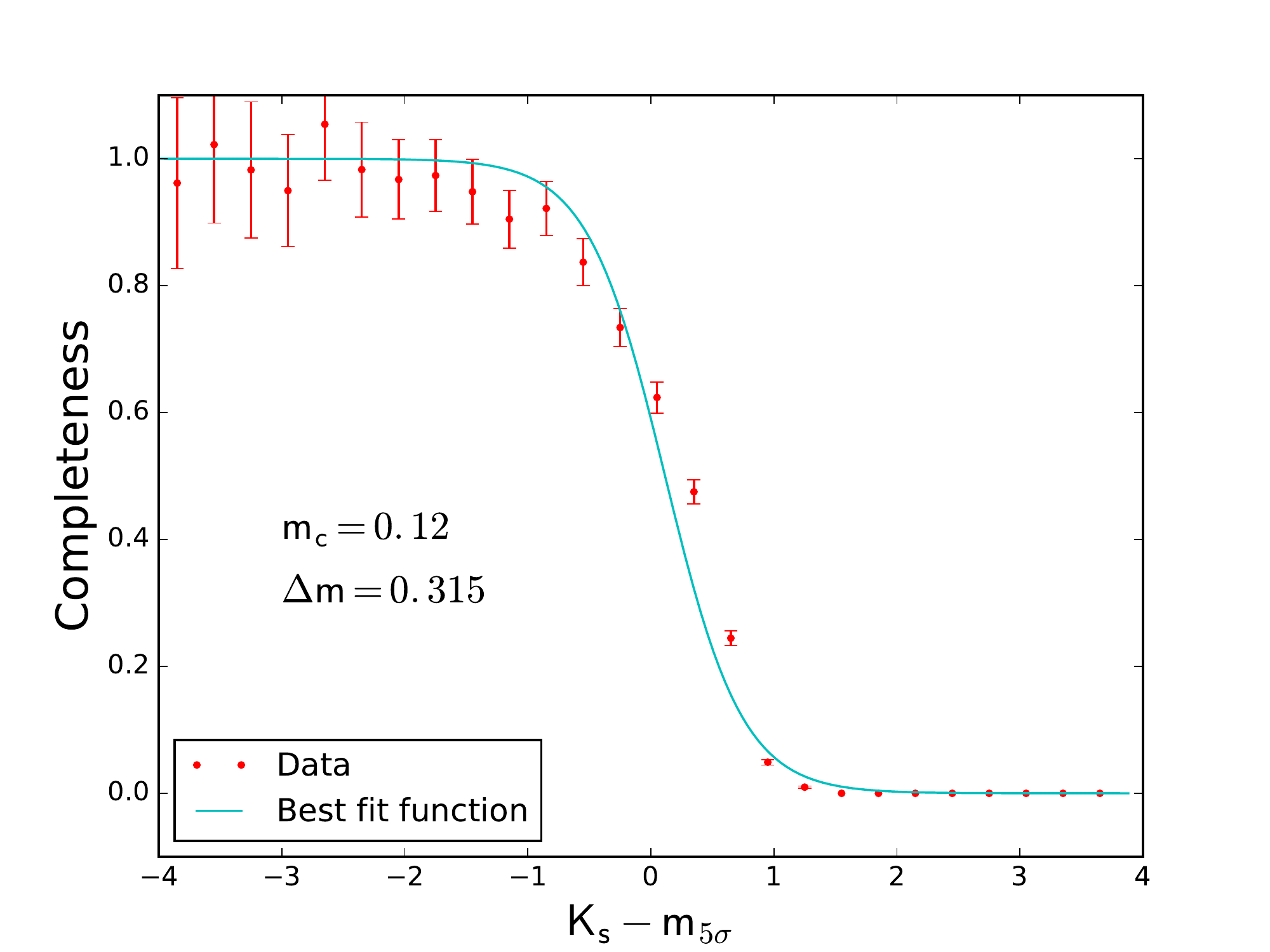}
\caption{Global completeness fraction for the ALHAMBRA-4 field as a function of the $K_s$ magnitude. The completeness has been calculated comparing our counts with those in \citet{2013ApJS..206....8M}, taking in account the UltraVISTA completeness function correction.} \label{completenessF}
\end{figure}

We repeated the fit a second time, in this case to the whole ALHAMBRA-4 field and substituting the parameter $m_{\rm{c}}$ for $(m_{\rm{c}}-m_{5\sigma}^{(i)}) \ (i=1,\dots,4)$, so that we obtain a single completeness function which can be applied to all four CCDs by plugging the value of the nominal $5\sigma$ detection limit in each CCD, thus "sliding" the global Fermi function to its adequate position. We plot in Figure \ref{completenessF} the completeness function thus derived. 

Our analysis also includes the correction by the completeness function of the UltraVISTA survey itself as presented by the authors, although at these magnitudes (which are bright compared to its limit) the correction is very small: the completeness fraction of the UltraVISTA catalogue is >90\% for magnitudes brighter than AB=23.4 \citep{2013ApJS..206....8M}.

 %%%%%%%%%
%%%%%%%%% 
 
We then use the full completeness function estimate to correct the ALHAMBRA $K_{s}$ band number counts. In order to obtain an accurate estimate of the number counts we have summed the number of objects in each magnitude interval, weighing each individual object both with the value of the completeness correction corresponding to its magnitude and with the accessible masked area in the particular CCD where it is observed. We have extended the number counts out to the 60\% completeness limit of each pointing, which allows us to reach the global completeness limit $K_s \approx 21.9$, as can be seen in Figure \ref{NumberCounts}. The global data are well fit with a power-law, and as expected we observe a reasonable behaviour of the number counts down to the catalogue limit. 
 
Finally, we have also checked that the results presented in this work are compatible with the NIR galaxy counts that were presented in \cite{2009ApJ...696.1554C} for the ALHAMBRA-8 field.

\section{The catalogue structure}
\label{appcolumns}

We list in this Appendix the items contained in our catalogue for each of the detected objects. We include more complete details on some of the items in the following paragraphs.

\begin{table}
\centering
\caption{Content and type of the columns in the catalogue files.}
\begin{tabular}{clc} 
\hline 
COLUMNS & CONTENT & TYPE \\ 
\hline
1       & ID Number$^a$ & Integer \\
2,3     & (X,Y) pixel coordinates & Real \\
4,5     & RA, Dec (J2000) & Real \\
6       & Area (pixels) & Integer \\
7,8     & F365W flux, error$^b$ & Real \\
9,10    & F365W magnitude, error$^b$ & Real \\
...     & ...   & ... \\
95,96   & $K_s$ flux, error$^b$ & Real \\
97,98   & $K_s$ magnitude, error$^b$ & Real \\
99,100  & Synth F814W flux, error$^b$ & Real \\
101,102 & Synth F814W magnitude, error$^b$ & Real \\
103     & SExtractor \texttt{FLAG} & Integer \\
104     & SExtractor \texttt{CLASS\_STAR} & Real \\
105     & \texttt{COLOUR\_CLASS\_STAR}$^c$ & Real \\
106     & \texttt{MASK\_SELECTION}$^d$ & Boolean \\
107     & BPZ photometric redshift & Real \\
108,109 & BPZ photo-z 95\% interval & Real \\
110     & BPZ SED type & Real \\
111     & BPZ {\it Odds} & Real \\
112     & BPZ stellar mass (log$_{10}$, $M_\odot$) & Real \\
113     & BPZ absolute $B_{\rm Johnson}$ & Real \\
114     & BPZ ML photo-z$^e$ & Real \\
115     & BPZ ML SED type$^e$ & Real \\
116     & BPZ fitting $\chi^2$ $^e$ & Real \\
117     & Absolute $K_s$ & Real \\ 
\hline
\end{tabular}
\end{table}

(a) The first column provides a unique ID for each source, built according to the following rule:

\begin{center}
$\mathrm{\underbrace{220}_{K_s-band} + \underbrace{1}_{Field}+\underbrace{1}_{Pointing}+\underbrace{1}_{CCD}+\underbrace{00001}_{SExtractor \quad ID}}$  
\end{center}

(b) All fluxes and magnitudes have been measured using the isophotal method in SExtractor. In those cases where the measured flux is less than its associated uncertainty, the magnitude value has been set to 99.0 and the magnitude error corresponds to the $1\sigma$ limit.

(c) Colour-based stellarity as defined in Section
\ref{stargal}.

(d) Indicates whether the object lies in the clean area after the mask described in Section \ref{masks} is applied.

(e) BPZ outputs the result of a pure maximum likelihood calculation of the photometric redshift, not including the type-luminosity-redshift Bayesian priors. We list in these columns such maximum likelihood-based best-fitting values, and the associated $\chi^2$ value.

%%%%%%%%%%%%%%%%%%%%%%%%%%%%%%%%%%%%%%%%%%%%%%%%%%
% Don't change these lines
\bsp	% typesetting comment
\label{lastpage}
\end{document}